\documentclass[twocolumn, twocolumnappendix]{aastex63}

\newcommand{\spi}{{\em Spitzer}}

\newcommand{\wise}{{\em WISE}}
\newcommand{\iras}{{\em IRAS}}
\newcommand{\iso}{{\em ISO}}

\newcommand{\mum}{\mbox{$\mu$m}}
\newcommand{\acet}{{\mbox{{C$_2$H$_2$}}}}
\newcommand{\smn}{{\mbox{[6]$-$[9]}}}
\newcommand\msun{\mbox{$\mathrm M_\odot$}}
\newcommand{\kmps}{km~s$^{-1}$}

\accepted{17 October 2019}
\submitjournal{ApJ}

\shorttitle{FORCAST Observations of Carbon Stars}
\shortauthors{Kraemer et al.}

\begin{document}

\title{Stellar Pulsation and the Production of Dust and Molecules in Galactic 
Carbon Stars}

\correspondingauthor{Kathleen E. Kraemer}
\email{kathleen.kraemer@bc.edu}

\author[0000-0002-2626-7155]{Kathleen E. Kraemer}
\affiliation{Institute for Scientific Research, Boston College,
140 Commonwealth Avenue, Chestnut Hill, MA 02467, USA;
kathleen.kraemer@bc.edu}

\author[0000-0003-4520-1044]{G. C. Sloan}
\affiliation{Space Telescope Science Institute, 3700 San Martin
Drive, Baltimore, MD 21218, USA; sloan@astro.cornell.edu}
\affiliation{Department of Physics and Astronomy, University of
  North Carolina, Chapel Hill, NC 27599-3255, USA}
\affiliation{Center for Astrophysics and Planetary Science, Cornell
  University, Ithaca, NY 14853-6801, USA}

\author[0000-0003-1046-512X]{Luke D. Keller} 
\affiliation{Department of Physics and Astronomy, Ithaca College}

\author[0000-0003-0356-0655]{Iain McDonald}
\affiliation{Jodrell Bank Centre for Astrophysics, Alan Turing Building, 
University of Manchester, Manchester, M13 9PL, UK}

\author[0000-0002-3171-5469]{Albert A. Zijlstra}
\affiliation{Jodrell Bank Centre for Astrophysics, Alan Turing Building, 
University of Manchester, Manchester, M13 9PL, UK}
\affiliation{Laboratory for Space Research, The University of Hong Kong, 
Pokfulam Road, Hong Kong}

\author[0000-0003-2723-6075]{Martin A. T. Groenewegen}
\affiliation{Koninklijke Sterrenwacht van Belgi\"e, Ringlaan 3, B--1180 
Brussels, Belgium}

\begin{abstract}
New infrared spectra of 33 Galactic carbon stars from FORCAST on SOFIA reveal 
strong connections between stellar pulsations and the dust and molecular
chemistry in their circumstellar shells.  A sharp boundary in overall dust
content, which predominantly measures the amount of amorphous carbon,
separates the semi-regular and Mira variables, with the
semi-regulars showing little dust in their spectra and the
Miras showing more.  In semi-regulars, the contribution from
SiC dust increases rapidly as the overall dust content grows,
but in Miras, the SiC dust feature grows weaker as more dust
is added.  A similar dichotomy is found with the absorption
band from CS at $\sim$7.3 $\mu$m, which is generally limited to
semi-regular variables.  Observationally, these differences make it
straightforward to distinguish semi-regular and Mira
variables spectroscopically  without the need for 
long-term photometric observations or knowledge of their distances. 
The rapid onset of strong SiC emission in Galactic carbon
stars in semi-regulars variables points to a different
dust-condensation process before strong pulsations take over.
The break in the production of amorphous carbon between
semi-regulars and Miras seen in the Galactic sample is also
evident in Magellanic carbon stars, linking strong pulsations
in carbon stars to the strong mass-loss rates which will end
their lives as stars across a wide range of metallicities.

\end{abstract}


\section{Introduction\label{sec.intro}}

A fundamental problem in astrophysics is understanding how stars in the 
late stages of their
evolution enrich galaxies with dust and freshly fused elements. It is
currently uncertain what the relative contributions are from the most massive
stars, which will explode as supernovae, creating new elements but 
likely destroying
dust, compared to lower-mass stars \citep[e.g.,][]{micelottaea18, dellagliea19,
nanniea19}. Low- and intermediate-mass stars create
dust grains in the cool outer layers of their atmospheres, and the radiation
pressure on these grains then helps drive the mass-loss process 
\citep[e.g.,][and references therein]{ho18, gs18}.
Dust serves as a readily observable tracer of the total mass loss,
and it is easily measured in the outflows from evolved stars. Infrared
surveys of the Large and Small Magellanic Clouds (LMC and SMC)
have pointed to carbon stars as the dominant source of dust being pumped into
the interstellar medium (ISM) by stars \citep[e.g.,][]{mbz09, 
boyerea11}.

A significant fraction of intermediate-mass stars may end their lives on the 
asymptotic
giant branch (AGB) as carbon stars, due to the production of carbon in their
interiors and its dredge-up to their surfaces. The AGB stars 
become carbon stars when their C/O ratios exceed unity. In 
these stars, free carbon
remains after the formation of CO, leading to a carbon-rich,
rather than oxygen-rich, gas and dust chemistry \citep[e.g.,][and references 
therein]{hab96,wk98}. The Infrared Spectrograph 
\cite[IRS;][]{irs04} on the {\em Spitzer Space Telescope} \citep{spitzer04}
has been used to study carbon stars in the low-metallicity environments of 
the LMC \citep[e.g.,][]{zijlstraea06,
matsuuraea06,leisenringea08}, the SMC \citep[e.g.,][]{smcc06,
lag07}, and other Local Group galaxies \citep[e.g.,][]{matsuuraea07,
sml12}. 

The key to those studies, particularly in the Magellanic Clouds, has been 
samples of sufficient size. The combined IRS programs used by \cite{mcc16}
had 144 carbon stars in the LMC and 40 in the SMC. 
Their control sample for the much larger
Milky Way, however, contained only 42 carbon stars.  The Galactic sample
was observed with the Short-Wavelength Spectrometer \citep[SWS;][]{sws96}
on the {\em Infrared Space Observatory} \citep[\iso;][]{iso96}
and served as a valuable comparison sample for many of the studies
of extragalactic carbon stars mentioned above. However, it is a 
 selective and relatively small sample given the size of the Milky Way. While
larger samples of infrared spectra were obtained with the Low-Resolution
Spectrometer \citep[LRS;][]{lrs86} on the {\em Infrared Astronomical 
Satellite} \citep[\iras;][]{iras84}, the shortest wavelength is 7.67 \mum,
too red to measure important spectral diagnostics.

To improve the Galactic sample, we obtained 4.9--13.7 
\mum\ spectra of Galactic carbon stars using
the Faint Object infraRed CAmera for the SOFIA Telescope 
\citep[FORCAST;][]{forcast} on the Stratospheric Observatory for Infrared
Astronomy \citep[SOFIA;][]{sofia}. Section \ref{sec.obs} describes our
source selection, observations, and data processing. The results are given 
in Section \ref{sec.res} and discussed
in Section \ref{sec.disc}, including comparisons to the SWS sample and
the Magellanic Cloud samples. Section \ref{sec.summ} summarizes our 
findings. The Appendix details the changes to the procedures for extracting
spectral features needed to avoid telluric contamination.

\section{Source Selection and Observations\label{sec.obs}}

\subsection{Source Selection\label{sec.srcs}}

\begin{figure}
\includegraphics[width=3in]{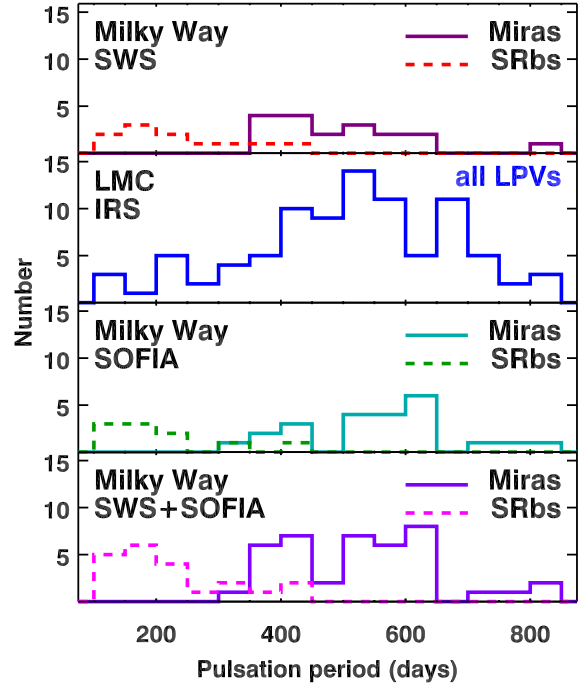}
\caption{Galactic and LMC carbon stars with 5--14 \mum\ spectra.
(top) Galactic stars with SWS spectra; (second from top) LMC long period 
variables (LPVs) with IRS spectra; (second from bottom) Galactic stars 
with new SOFIA spectra;
(bottom) combined SWS and SOFIA sample of Galactic stars. The Galactic samples
are separated into Miras (solid lines) and SRbs (dashed lines).
}
\label{fig.bias}
\end{figure}

The Galactic sample from the SWS consists of observations from 
numerous individual projects and suffers from 
inevitable biases due to the various selection criteria used. 
In particular, the SWS sample has a deficit of semi-regular variables,
which pulsate with weaker amplitudes than Miras. The SWS sample also
selected against carbon stars with longer pulsation periods,
especially $P >$ 400 d, compared to the Magellanic samples
(Fig. \ref{fig.bias}, top two panels). These sources are the most embedded 
objects with the
most optically thick dust shells, and as noted above, they may dominate the 
dust returned to the ISM by stars.

To address the biases in the SWS data, we started with samples of 
Galactic carbon stars from \cite{slmp98}, \cite{jonesea90}, 
\cite{nakashimaea00}, and \cite{wfm06}. 
Using their periods and the \iras\ photometry, we selected sources in three 
groups. Group 1 has SRb variables with $F_\nu$ (12 \mum) $>$ 40 Jy; Group 2 
contains Mira variables with 400 d $< P <$ 500 d and $F_\nu >$ 
60 Jy; and Group 3 consists of Miras with $P >$ 550 d and F$_\nu$ 
$>$ 150 Jy\footnote{The flux constraints were driven by the desire for short 
integration
times and a reasonable sample size.}. Our sample originally had 22 stars in 
Group 1, 8 in Group 2, and 13 in Group 3. Due to scheduling constraints 
that necessitated using back-up targets, as well as 
the nature of SOFIA Survey Programs, the observed sample
consists of 11 Group 1 stars, 6 Group 2 stars, and 18 Group 3 stars, which
yielded 33 usable spectra.

\subsection{FORCAST Observations and Pipeline Processing}

SOFIA observed the stars in service mode over the course of three 
cycles, Cycle 1 (2013, 4 stars, Plan ID=01\_0041), Cycle 3 (2015, 10 stars,
03\_0104), and Cycle 4 (2016, 21 stars, 04\_0129). We used the two 
low-resolution grism settings in FORCAST, G063 and
G111, to obtain spectra over the wavelength ranges 4.9--7.8 and 8.3--13.8 \mum,
respectively \cite{kellerea10}. The observations were made with the 
4\farcs7 slit, which results
in a spectral resolution of R $\sim$ 120. The 31 stars from Cycles 3 and 4 were
processed by the SOFIA Data Cycle System with pipeline version 1\_2\_0.
Those from Cycle 1 were processed with pipeline version 1\_0\_0; they could
not be re-processed due to changes in the observing configuration 
after Cycle 1. Two sources (CIT 6 and T Cnc) were discarded as their data 
were inconsistent both with previous observations and between the G063 and 
G111 spectra. Thus, our final sample consists of 33 stars. The bottom two
panels in Figure \ref{fig.bias} show our new SOFIA sample and the combined
sample for the Milky Way.

\begin{deluxetable*}{lccccrcr}
\tablecaption{FORCAST Targets\label{tab.tgts}}
\tablewidth{0pt}
\tablehead{
\colhead{Star} & \colhead{RA} & \colhead{Dec}  &
\colhead{Obs. Date} & \colhead{Period} & \colhead{$F_{12}$} &
  \colhead{} & \colhead{Period} \\
\colhead{Name} & \multicolumn{2}{c}{(J2000.0)} & 
\colhead{(yyyy-mm-dd)} &  \colhead{(days)} 
& \colhead{(Jy)} &  \colhead{Group}
& \colhead{Ref.} 
}
\startdata
WZ Cas     & 00 01 15.86& +60 21 19.0   & 2016-09-21 & 186 &  44   & 1 & S98\\
ST Cam     & 04 51 13.35& +68 10 07.6   & 2013-09-13 & 300 &  95   & 1 & S98\\
Y Tau      & 05 45 39.41& +20 41 42.2   & 2016-09-22 & 242 & 144   & 1 & S98\\
TU Gem     & 06 10 53.10& +26 00 53.4   & 2015-11-10 & 230 &  70   & 1 & S98\\
UU Aur     & 06 36 32.84& +38 26 43.8   & 2013-09-13 & 234 & 232   & 1 & S98\\
X Cnc      & 08 55 22.88& +17 13 52.6   & 2015-05-30 & 195 &  90   & 1 & S98\\
T Cnc\tablenotemark{a} & 08 56 40.15& +19 50 57.0 & 2015-06-05 & 482 &  61   & 1 & S98\\
U Hya      & 10 37 33.27& $-$13 23 04.3 & 2015-06-03 & 450 & 206   & 1 & S98\\
TW Oph     & 17 29 43.66& $-$19 28 22.9 & 2016-02-18 & 185 &  96   & 1 & S98\\
RT Cap     & 20 17 06.53& $-$21 19 04.5 & 2016-07-14 & 393 &  73   & 1 & S98\\
RV Cyg     & 21 43 16.33& +38 01 03.0   & 2015-05-30 & 263 & 103   & 1 & S98\\
KY Cam     & 03 27 59.03& +60 44 55.2   & 2016-09-27 & 477: &  70   & 2 & N00\\
V718 Tau   & 04 31 21.93& +17 39 10.3   & 2015-09-18 & 405 &  72   & 2 & S98\\
R Lep      & 04 59 36.35& $-$14 48 22.5 & 2016-02-10 & 427 & 380   & 2 & S98\\
CL Mon     & 06 55 36.69& +06 22 43.2   & 2015-11-06 & 497 & 113   & 2 & S98\\
U Cyg      & 20 19 36.59& +47 53 39.1   & 2015-06-04 & 463 & 112   & 2 & S98\\
AX Cep     & 21 26 54.03& +70 13 15.4   & 2016-09-21 & 395 &  72   & 2 & S98\\
V668 Cas   & 00 27 41.13& +69 38 51.6   & 2016-09-20 & 650 & 306   & 3 & J90\\
V370 Aur   & 05 43 49.69& +32 42 06.2   & 2016-02-05 & 780 & 196   & 3 & J90\\
V1259 Ori  & 06 04 00.05& +07 25 52.0   & 2016-02-05 & 686 & 320   & 3 & J90\\
CGCS 6276  & 08 09 20.26& $-$36 24 26.8 & 2016-02-05 & 832 & 156   & 3 & W06\\
V346 Pup   & 08 10 48.89& $-$32 52 05.6 & 2016-02-05 & 568 & 347   & 3 & W06\\
CQ Pyx     & 09 13 53.94& $-$24 51 25.2 & 2016-02-05 & 659 & 737   & 3 & W06\\
IRC +10216 & 09 47 57.41& +13 16 43.6   & 2015-11-20 & 651 &47530  & 3 & W06\\
CGCS 2653  & 09 53 06.72& $-$53 38 53.5 & 2016-07-12 & 630 & 157   & 3 & W06\\
CIT 6\tablenotemark{a} & 10 16 02.28& +30 34 19.0 & 2013-10-25 & 617 & 3319  & 3 & W06\\
CGCS 2987  & 11 16 38.90& $-$65 50 56.0 & 2016-07-18 & 623 & 181   & 3 & W06\\
V1132 Cen  & 12 42 09.58& $-$43 55 02.9 & 2016-07-18 & 551 & 159   & 3 & W06\\
CGCS 3311  & 12 57 15.79& $-$69 01 51.5 & 2016-07-14 & 586 & 264   & 3 & W06\\
V2548 Oph  & 17 07 58.11& $-$24 44 31.2 & 2016-02-18 & 747 & 793   & 3 & W06\\
RAFGL 2154 & 18 26 39.48& $-$06 54 03.6 & 2016-02-11 & 635 & 220   & 3 & W06\\
CGCS 4014  & 18 27 34.22& $-$08 37 23.4 & 2016-09-21 & 659 & 153   & 3 & W06\\
V1420 Aql  & 19 20 18.12& $-$08 02 12.1 & 2013-09-19 & 694 & 384   & 3 & W06\\
V1969 Cyg  & 20 09 14.24& +31 25 44.9   & 2015-11-20 & 550 & 173   & 3 & J90\\
V384 Cep   & 22 25 53.47& +60 20 43.6   & 2016-09-21 & 698: & 182   & 3 & N00\\
\enddata
\tablenotetext{a}{Observations problematic and not included in analysis.}
\tablereferences{J90: \cite{jonesea90}; S98: \cite{slmp98, gcvs4}; N00: 
\cite{nakashimaea00}; W06: \cite{wfm06}.}
\end{deluxetable*}

\subsection{Comparison Samples}

The comparison sample for the Milky Way comes from the SWS archive of 
\cite{swsatlas}, and carbon-rich sources were selected using the spectral 
classifications by \cite{kspw02}. We used the variability types and
periods from the General Catalog of Variable Stars \cite[GCVS;][]{gcvs17}
to assign 31 of the 42 sources to one of the three SOFIA-selection groups if 
possible\footnote{We dropped the flux criterion (Sec. 2.1). Also, periods for 
V Aql and Y 
CVn have been updated since \cite{mcc16} to 400 and 268 d, respectively.};
two Miras with periods of 358 and 389 d were placed in Group 2.
Of the 11 SWS sources that did not fit in these groups, 3 are SRa 
variables (2 of which have known periods), 4 were Lb variables (i.e., 
irregulars), and 4 
have no variability type assigned in the GCVS.

We use the Magellanic Cloud sample from \cite{mcc16}, who compiled and
uniformly processed IRS observations of carbon stars from 11 \spi\ programs.
Although the Magellanic sample does come from several separate programs,
some of those projects were large surveys that attempted to sample a broad
range of sources. Thus, the Magellanic sample, though not completely free
of selection biases, is nonetheless less biased than the SWS sample.

As with the Galactic samples, we assign the Magellanic sources to one of 
the three 
groups, using variability types and periods from \cite{sus09, sus11} for the
LMC and SMC, respectively. These are based on data from the Optical 
Gravitational Lensing Experiment \cite[OGLE;][]{ogle92}. For the LMC,
55 of the 144 sources could be assigned a group, and for the SMC 32 of 40. 
We have left Magellanic Miras with periods below 395 d unassigned for better
comparison with the Galactic sample. OGLE does not distinguish among
different semi-regular categories. We have treated them as SRa or SRb
depending on their position in period-luminosity space (see Sec. 
\ref{sec.mode}): 4 of the 13 SRs in the LMC sample are SRbs and 3 of the 
12 SMC SRs are SRbs.

\section{Results and Analysis\label{sec.res}}

\subsection{The Spectra}
\begin{figure*}
\includegraphics[width=6.7in]{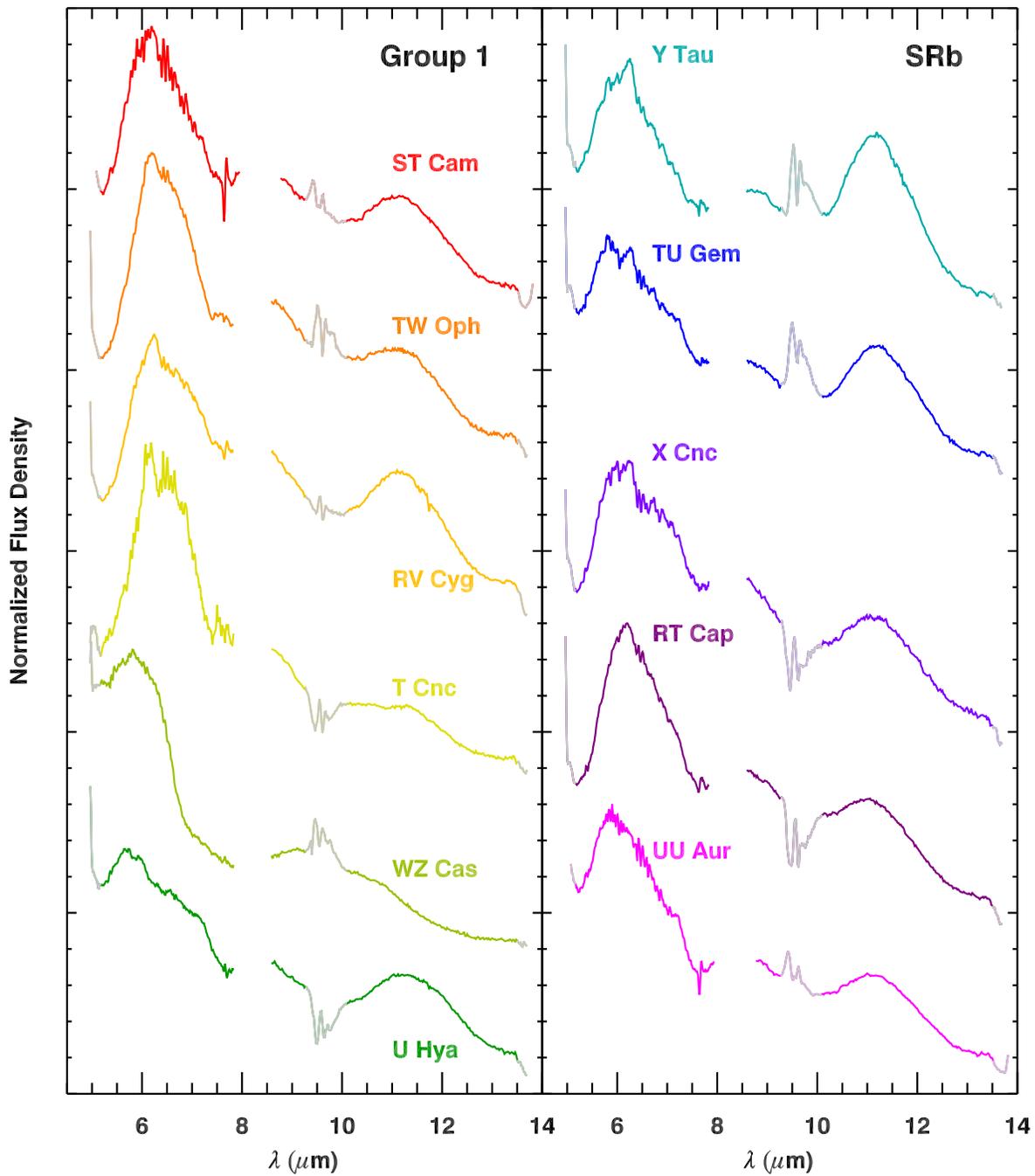}
\caption{Group 1:{ SRb variables, $F_\nu$ (12 \mum) $>$ 40 Jy}. 
Segments in 
gray are contaminated by telluric ozone (9.25 to 10.1 \mum) or instrumental
artifacts ($<$5.17 and $>$13.5 \mum). A telluric residual is also
sometimes present at 7.6--7.7 \micron. Spectra have been normalized and offset for clarity.
}
\label{fig.gr1}
\end{figure*}

\begin{figure}
\includegraphics[width=3.4in]{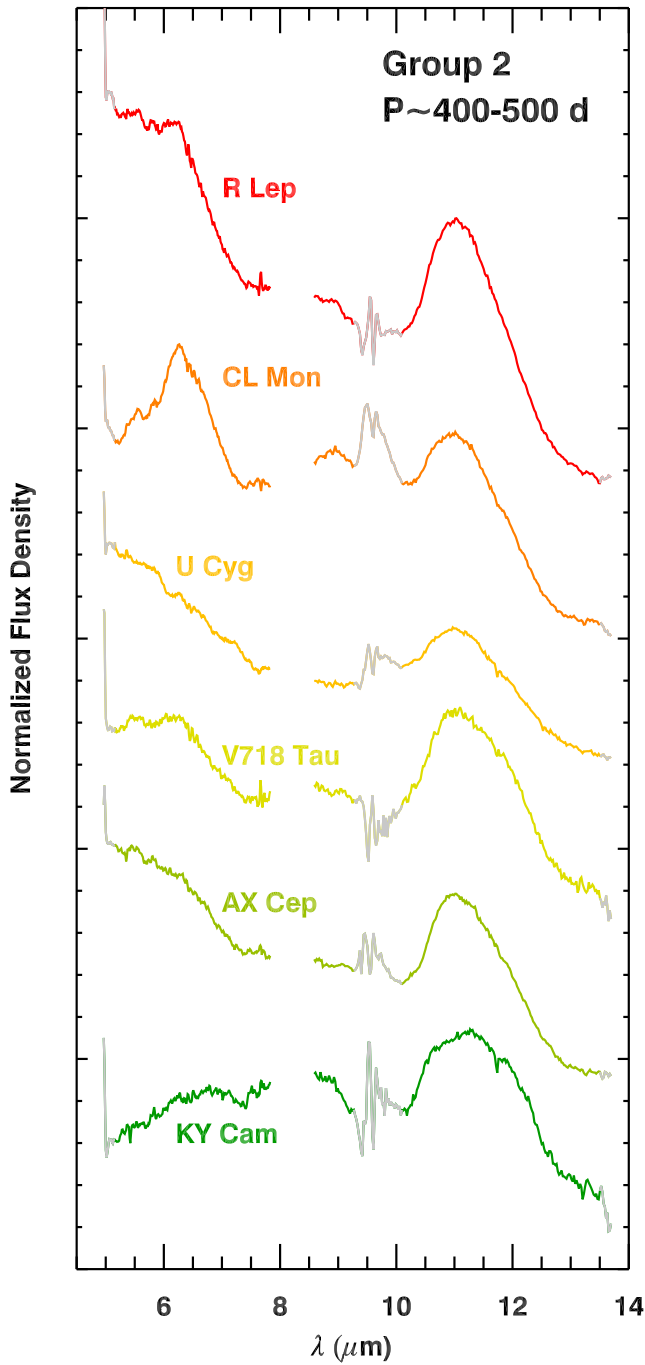}
\caption{Group 2: { Mira variables with 400 d $< P <$ 500 d and $F_\nu >$ 
60 Jy.}}
\label{fig.gr2}
\end{figure}

\begin{figure*}
\includegraphics[width=6.7in]{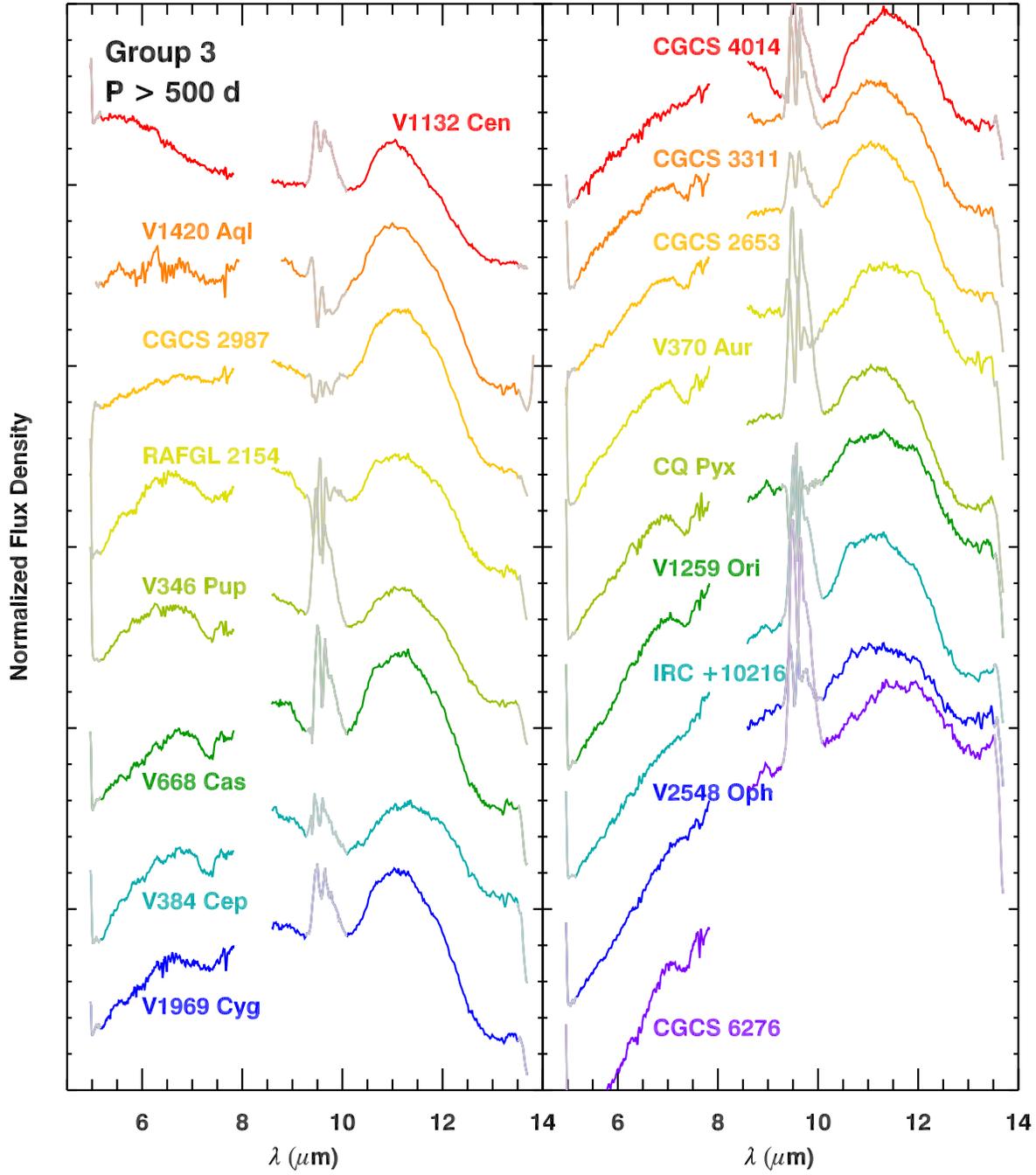}
\caption{Group 3:{ Miras with $P >$ 550 d and F$_\nu$ 
$>$ 150 Jy.}}

\label{fig.gr3}
\end{figure*}

The new FORCAST spectra for most of the Galactic carbon stars are 
dominated by the SiC dust emission feature at $\sim$11 \mum\ in the G111 
spectral range, while the G063 data show absorption features from 
carbonaceous molecules such as CO, \acet\, and CS.
Figures \ref{fig.gr1}--\ref{fig.gr3} show the data for Groups 
1--3, respectively. The 
wavelength range from 9.25 and 10.1 \mum, shown in gray, is contaminated 
by telluric ozone that was incompletely removed due to 
changes in the atmospheric absorption between observations of science and 
calibration 
targets. The range
from 7.6 to 7.7 \mum\ can also show telluric contamination in some of the
spectra (e.g., ST Cam). An instrumental artifact is present in some spectra, 
causing a spike below $\sim$5.2 \mum\ and a drop above 13.5 
\mum\footnote{https://www.sofia.usra.edu/sites/default/files/Uspot\_
{DCS\_DPS/Documents/}Known\_Issues\_010518.pdf}.
These data are excluded from further analysis, so the
\acet\ absorption feature at 13.7 \mum\ was not considered.

\subsection{Spectral Feature Extraction}

\begin{figure}
\includegraphics[width=3.25in]{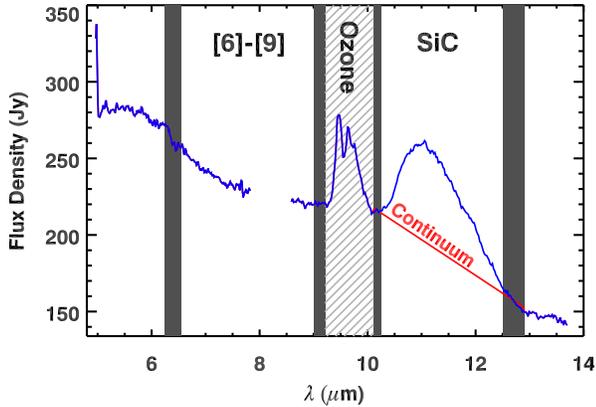}
\caption{Feature extraction example with V1132 Cen. The \smn\ color is 
measured from the 
leftmost gray bands. The silicon carbide feature using the rightmost gray
bands and the red line for the continuum level. The gray-striped region
is contaminated by telluric ozone.}
\label{fig.ext}
\end{figure}

\begin{deluxetable*}{rccccc}
\tablecaption{Dust Properties\label{tab.dust}}
\tablewidth{0pt}
\tablehead{
\colhead{Star} & \colhead{\smn} & \colhead{$F_{SiC}$/}  &
\colhead{$\lambda_C$} & \colhead{log $\dot{D}$} & \colhead{CE}\\
\colhead{Name} & \colhead{(mag)} &\colhead{Continuum} & 
\colhead{(\mum)} & \colhead{(\msun/yr)}& \colhead{Class}
}
\startdata
\multicolumn{6}{c}{Group 1}\\\hline
    WZ Cas & 0.069 $\pm$ 0.015 &$-$0.021 $\pm$ 0.005 & 11.98 $\pm$ 0.18 & $-$8.79 & 1\\
    ST Cam & 0.180 $\pm$ 0.010 & 0.187 $\pm$ 0.009 &  11.30 $\pm$ 0.07 & $-$8.61 & 1\\
     Y Tau & 0.287 $\pm$ 0.014 & 0.368 $\pm$ 0.004 &  11.25 $\pm$ 0.01 & $-$8.44 & 1\\
    TU Gem & 0.281 $\pm$ 0.010 & 0.266 $\pm$ 0.004 &  11.29 $\pm$ 0.02 & $-$8.45 & 1\\
    UU Aur & 0.228 $\pm$ 0.008 & 0.153 $\pm$ 0.010 &  11.29 $\pm$ 0.09 & $-$8.54 & 1\\
     X Cnc & 0.269 $\pm$ 0.012 & 0.192 $\pm$ 0.003 &  11.28 $\pm$ 0.02 & $-$8.47 & 1\\
     U Hya & 0.285 $\pm$ 0.006 & 0.206 $\pm$ 0.005 &  11.37 $\pm$ 0.04 & $-$8.44 & 1\\
    TW Oph & 0.100 $\pm$ 0.008 & 0.173 $\pm$ 0.003 &  11.30 $\pm$ 0.02 & $-$8.74 & 1\\
    RT Cap & 0.178 $\pm$ 0.010 & 0.174 $\pm$ 0.002 &  11.24 $\pm$ 0.01 & $-$8.62 & 1\\
    RV Cyg & 0.252 $\pm$ 0.010 & 0.233 $\pm$ 0.003 &  11.28 $\pm$ 0.02 & $-$8.50 & 1\\\hline
\multicolumn{6}{c}{Group 2}\\\hline
    KY Cam & 0.754 $\pm$ 0.007 & 0.215 $\pm$ 0.005 &  11.31 $\pm$ 0.04 & $-$7.69 & 3\\
  V718 Tau & 0.500 $\pm$ 0.007 & 0.363 $\pm$ 0.007 &  11.27 $\pm$ 0.03 & $-$8.10 & 2\\
     R Lep & 0.316 $\pm$ 0.008 & 0.347 $\pm$ 0.004 &  11.19 $\pm$ 0.01 & $-$8.40 & 1\\
    CL Mon & 0.471 $\pm$ 0.007 & 0.239 $\pm$ 0.002 &  11.20 $\pm$ 0.01 & $-$8.15 & 2\\
     U Cyg & 0.433 $\pm$ 0.005 & 0.169 $\pm$ 0.002 &  11.22 $\pm$ 0.02 & $-$8.21 & 2\\
    AX Cep & 0.448 $\pm$ 0.005 & 0.347 $\pm$ 0.006 &  11.22 $\pm$ 0.02 & $-$8.18 & 2\\
\hline
\multicolumn{6}{c}{Group 3}\\\hline
  V668 Cas & 0.835 $\pm$ 0.006 & 0.207 $\pm$ 0.003 &  11.28 $\pm$ 0.02 & $-$7.56 & 3\\
  V370 Aur & 1.054 $\pm$ 0.003 & 0.115 $\pm$ 0.004 &  11.42 $\pm$ 0.06 & $-$7.21 & 4\\
 V1259 Ori & 1.263 $\pm$ 0.006 & 0.104 $\pm$ 0.003 &  11.37 $\pm$ 0.05 & $-$6.88 & 5\\
 CGCS 6276 & 1.526 $\pm$ 0.008 & 0.062 $\pm$ 0.003 &  11.50 $\pm$ 0.08 & $-$6.46 & 3\\
  V346 Pup & 0.765 $\pm$ 0.004 & 0.161 $\pm$ 0.003 &  11.34 $\pm$ 0.03 & $-$7.68 & 3\\
    CQ Pyx & 1.139 $\pm$ 0.004 & 0.130 $\pm$ 0.003 &  11.33 $\pm$ 0.03 & $-$7.08 & 4\\
IRC +10216 & 1.265 $\pm$ 0.006 & 0.146 $\pm$ 0.003 &  11.31 $\pm$ 0.03 & $-$6.88 & 5\\
 CGCS 2653 & 1.044 $\pm$ 0.002 & 0.157 $\pm$ 0.003 &  11.30 $\pm$ 0.03 & $-$7.23 & 4\\
 CGCS 2987 & 0.794 $\pm$ 0.002 & 0.200 $\pm$ 0.004 &  11.31 $\pm$ 0.03 & $-$7.63 & 3\\
 V1132 Cen & 0.577 $\pm$ 0.005 & 0.200 $\pm$ 0.002 &  11.21 $\pm$ 0.02 & $-$7.98 & 2\\
 CGCS 3311 & 1.039 $\pm$ 0.003 & 0.122 $\pm$ 0.003 &  11.34 $\pm$ 0.03 & $-$7.25 & 4\\
 V2548 Oph & 1.369 $\pm$ 0.007 & 0.084 $\pm$ 0.002 &  11.32 $\pm$ 0.04 & $-$6.71 & 5\\
RAFGL 2154 & 0.738 $\pm$ 0.006 & 0.169 $\pm$ 0.003 &  11.32 $\pm$ 0.03 & $-$7.72 & 3\\
 CGCS 4014 & 0.916 $\pm$ 0.006 & 0.177 $\pm$ 0.004 &  11.36 $\pm$ 0.04 & $-$7.43 & 3\\
 V1420 Aql & 0.740 $\pm$ 0.007 & 0.212 $\pm$ 0.011 &  11.30 $\pm$ 0.08 & $-$7.72 & 3\\
 V1969 Cyg & 0.865 $\pm$ 0.004 & 0.194 $\pm$ 0.003 &  11.29 $\pm$ 0.02 & $-$7.52 & 3\\
  V384 Cep & 0.871 $\pm$ 0.004 & 0.116 $\pm$ 0.003 &  11.40 $\pm$ 0.04 & $-$7.51 & 3\\
\enddata
\tablecomments{Uncertainties are 1$\sigma$
and primarily reflect the observational noise in the spectra.}
\end{deluxetable*}

We use the Manchester system \citep{smcc06,zijlstraea06} to measure
 the \smn\ color  and the strength of the silicon carbide dust feature (SiC)
relative to the local continuum (Fig. \ref{fig.ext}). The \smn\ color
measures the relative strengths of the star and the dominant dust component,
amorphous carbon, which has an  opacity that is a
smooth function of wavelength with no known resonances 
\citep[e.g.,][]{jura83a, jura86,mr87}.  The color  serves as a proxy for the 
dust-production rate 
\citep{groenewegenea07} and  is determined using two regions
of the spectra which are free of molecular bands and solid-state features.

Due to residuals from the 
telluric ozone feature in the SOFIA spectra, the wavelengths used to define
these features had to be adjusted compared to those used by \cite{mcc16}.
In order to compare 
apples to apples, we re-extracted the spectral features from the 
Magellanic (IRS) and Galactic (SWS) samples using the revised wavelength 
ranges. Appendix \ref{app.lambdas} gives the new values,
along the updated wavelength ranges. Although the exact values for a given star
have shifted slightly compared to \cite{mcc16}, the overall characteristics 
of the samples remain the same. Thus, we will refer to the color as \smn\ for
simplicity when comparing to values in the literature.

Table \ref{tab.dust}
gives the results for the \smn\ color and the relative SiC strength for the 
SOFIA sample. 
The parameter $\lambda_C$ is the wavelength at which
half of the flux in the SiC feature is to the red and half to the blue.
The table includes the infrared spectral classification by \cite{mcc16}, 
which is based on the \smn\ color and the
dust production rate (DPR or $\dot{D}$). $\dot{D}$  is calculated from the 
\smn\ color, 
assuming an outflow velocity $v_{{\rm out}}$ of 10 \kmps\ according to the 
formula\footnote{Incorrect
equations were given for the dust mass-loss rate by
\cite{zoo08} and \cite{sml12}. { The formula assumes optical constants
from \cite{rm91}; using those from \cite{zubkoea04} would give different 
values.}}

\begin{eqnarray}
  \nonumber \log~\dot{D} \left( \frac{M_{\odot}}{{\rm yr}} \right)=
  \log\left( \frac{v_{{\rm out}}}{10 \, {\rm km~ s}^{-1}} \right) \nonumber \\
  -\!8.9 + 1.6 \, ([6]-[9]).
\end{eqnarray}

\begin{figure}
\includegraphics[width=3.35in]{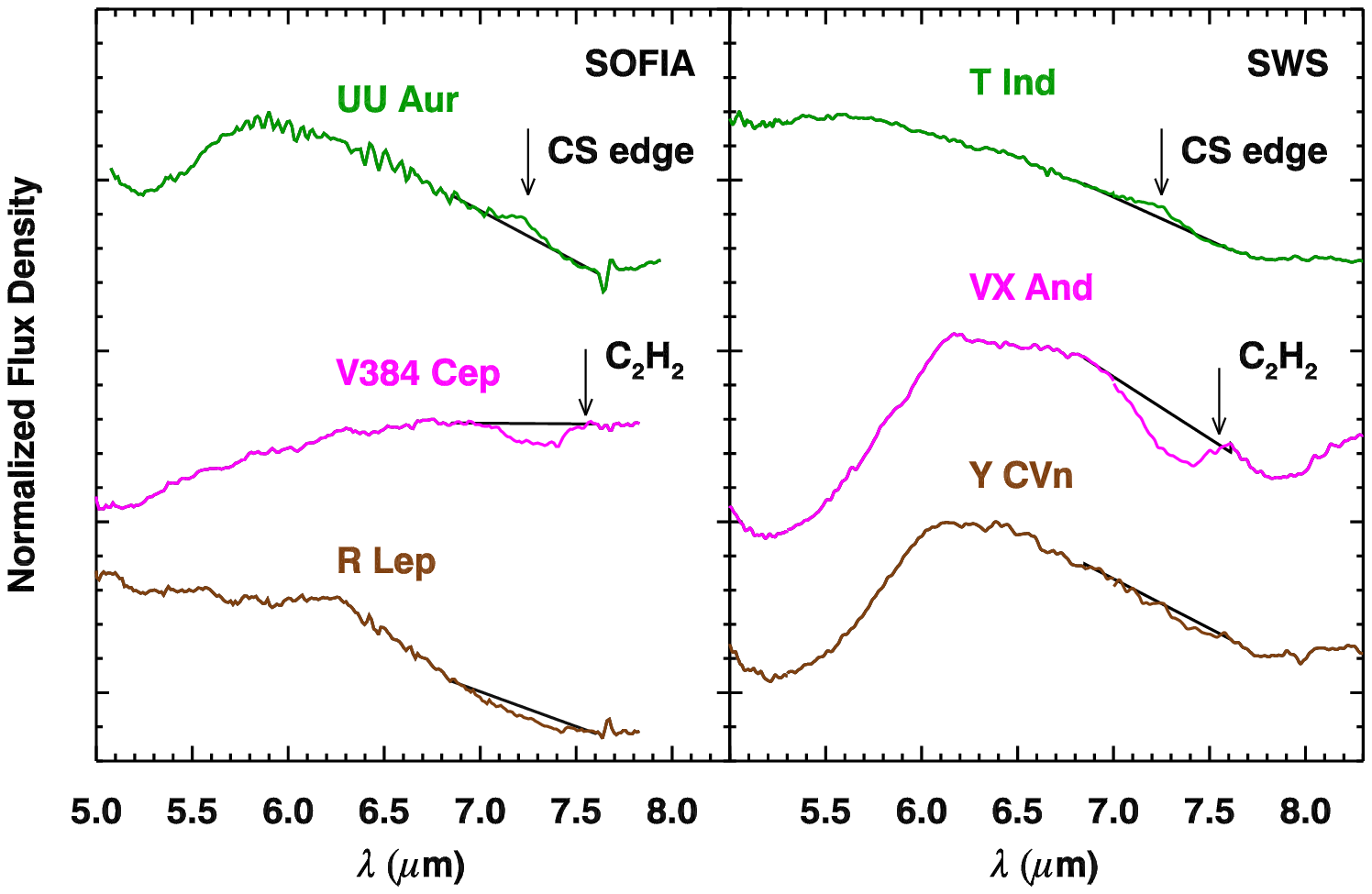}
\caption{The 6.5--8 \micron\ absorption complex. (left) Examples from the 
SOFIA data with (top) the CS absorption band, (middle) the 
\acet\ band, and
(bottom)  neither band apparent; 
(right) SWS examples.
Arrows indicate the features assessed by eye; the black lines
show the feature extraction region.
\label{fig.abs}
}
\end{figure}
 
We also can identify the presence of either \acet\ or CS absorption bands in 
many 
of the spectra between $\sim$7 and 7.6 \mum\ (rarely both at the same time). 
The $v_4+v_5$ \acet\ bands can be recognized by eye from the central peak of 
classic ``W'' shape at $\sim$7.5 \mum,  while a ``knee'' in the spectra reveals
the CS $\Delta v$ = 1 bandhead at $\sim$7.3 \mum. To 
characterize these absorption bands, we use a procedure similar to the 
Manchester method, with the blue end of the extraction 
set at 6.9 \micron\ and the red end at 7.57 \micron. These wavelengths
cover only half of the \acet\ band, but they avoid the telluric 
residual at 7.6--7.7 \micron\ and gap in coverage beyond 7.8 \mum.
Thus, we determine a partial equivalent width, EW$_p$, and
the wavelength centroid for the partial band, $\lambda_{p}$. We stress that
EW$_p$ and $\lambda_{p}$ are used here only as diagnostics because they only 
cover part of the bands. We apply the same wavelength stops defined for the 
FORCAST data to the SWS data as well. 

Figure \ref{fig.abs} (left) shows examples of FORCAST spectra that have 
absorption either from CS or from \acet, as well as a 
spectrum without a strong feature from either molecule (although 
absorption is clearly present in the general wavelength
region). The right-hand panel shows similar spectra from the SWS sample, with
the CS, and especially the \acet\ feature, more fully sampled. 
Many sources also show absorption below 6 \micron, likely from CO and/or
C$_3$ \citep[e.g.,][]{jorgensenea00}, but the limited wavelength coverage
prevents further analysis with the FORCAST spectra here.

\subsection{SiC Contrast and \smn\ Color}

\begin{figure}
\includegraphics[width=3.35in]{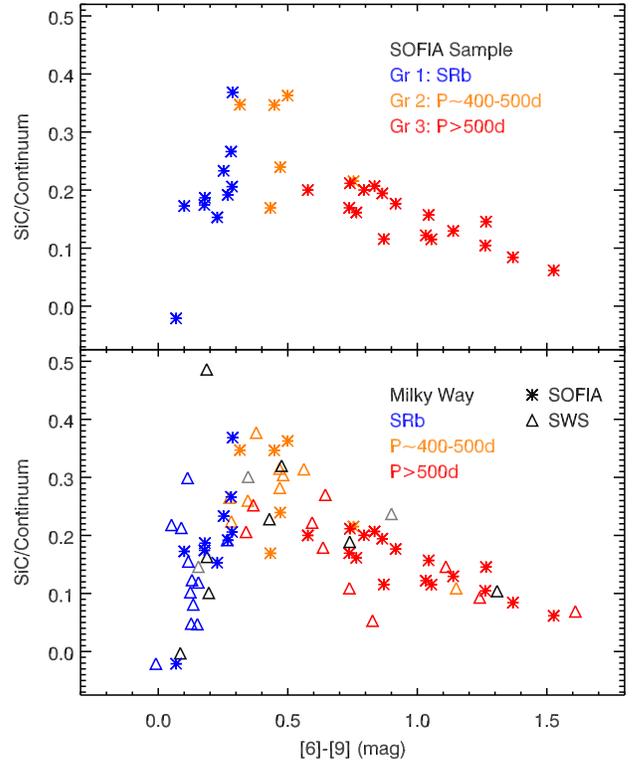}
\caption{\smn\ vs.\ SiC strength for the Milky Way. (top) The SOFIA sample, 
with colors corresponding to the selection group.
(bottom) The SWS sample added,
also colored by group (gray triangles are SRas, black triangles are 
Lbs or unknown). }
\label{fig.69silc_mw}
\end{figure}

\begin{figure}
\includegraphics[width=3.35in]{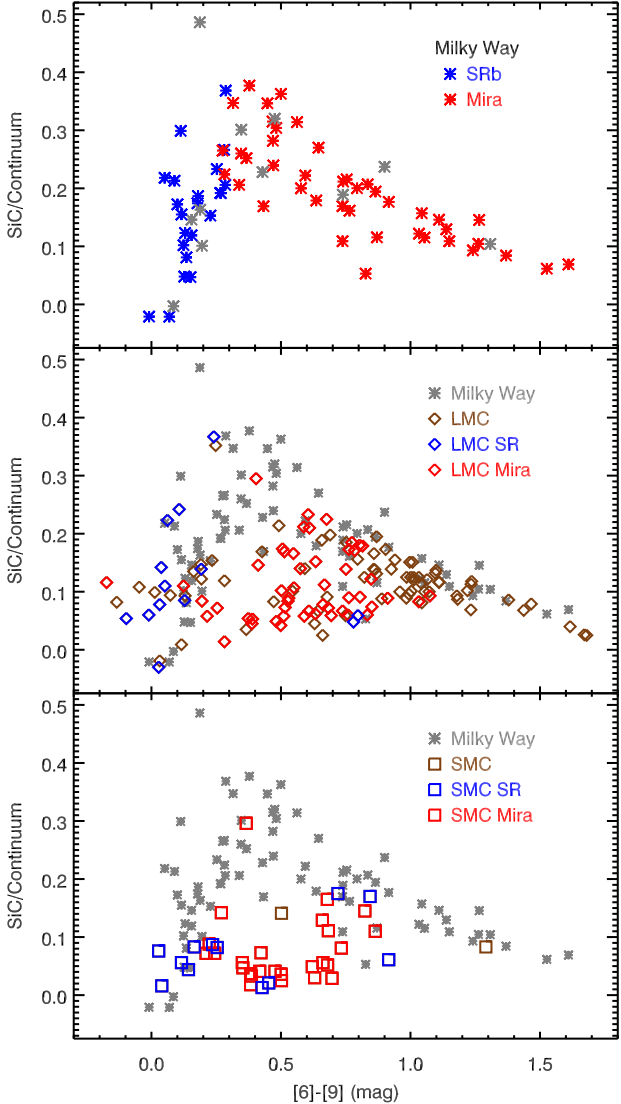}
\caption{\smn\ vs.\ SiC strength. (top) The combined Galactic sample, with 
SRbs in blue and Miras in red; { gray asterisks are Galactic SRas, Lbs, or 
unknown}. (middle) The LMC sample; 
sources with no OGLE class are in brown; Galactic sources are gray asterisks.
(bottom) The SMC sample, same colors as for the LMC. 
}
\label{fig.69silc_mc2}
\end{figure}

Figure \ref{fig.69silc_mw} shows the ratio of the SiC dust 
feature to the continuum as a function of \smn\ color, which traces 
the amorphous carbon content, with the SOFIA sample in the top panel.
A boundary at \smn\ $\sim$ 0.3 cleanly separates the  SRb
variables (Group 1) and Miras (Groups 2 and 3).  In the SRbs to the blue
of that boundary, the relative strength of the SiC dust feature rises 
steeply with increasing color, while in the Miras to the red, it
decreases more gradually as the color grows redder, with the
longer-period Miras associated with redder \smn\ colors.

The lower panel in Figure \ref{fig.69silc_mw} adds the SWS 
sample of Galactic carbon stars, colored by group as with the
SOFIA sample.  The boundaries and differences in behavior of
the SRbs and Miras remain unchanged.

Figure \ref{fig.69silc_mc2} compares the Galactic sample to 
those in the LMC and the SMC. In the lower two panels, sources not identified 
as semi-regular or Mira variables in the OGLE-III survey
\citep{sus09,sus11} appear as brown symbols, and the Galactic 
carbon stars are plotted in gray for comparison.  (LMC 
sources with \smn\ $>$ 1.8 are omitted here for clarity but 
are included in the table in the appendix.)

For the LMC, all but two of the 13 semi-regulars in the OGLE 
catalogs are on the same steeply rising segment as the 
Galactic SRb variables.  The LMC Miras are split between the 
segment with the Galactic Miras and a lower segment with weaker 
SiC/continuum ratios for the same \smn\ color.  The two 
segments rejoin at \smn\ $\sim$ 1.1.  The SMC sources generally
follow the lower segment with weaker SiC/continuum ratios.

\subsection{Molecular Gas Absorption}

\begin{figure}
\includegraphics[width=3.35in]{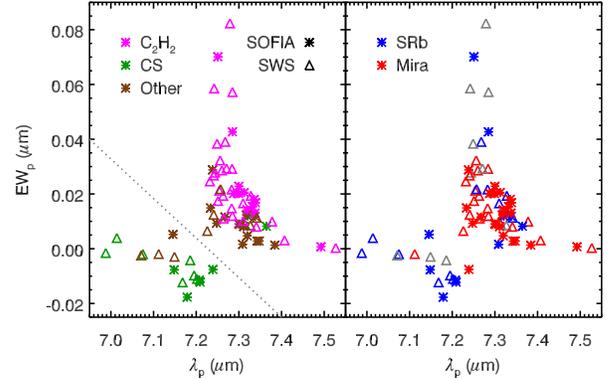}
\caption{Partial equivalent width vs. feature centroid. (left) Sources 
color coded by the qualitative assessment.
Almost all the CS sources (green) are in the lower left quadrant and the 
\acet\ sources (magenta) are in the upper right quadrant.
(right) Sources color coded by variability. The Miras
are almost entirely in the \acet\ region, while the SRbs are split between
CS and \acet. The dashed line in the left panel is simply to guide the eye; 
{ gray symbols are the SRas, Lbs, and unknowns, as in Fig. 8}.
\label{fig.ew7}
}
\end{figure}

The G063 grism of FORCAST covers the 4.9--7.8 \mum\ 
range, which for carbon stars can include absorption bands 
from carbon-bearing molecules such as \acet, HCN, CS, C$_3$, 
and CO \citep[e.g.,][and references therein]{goebelea80, 
goebelea81,aokiea98, aokiea99, jorgensenea00,speckea06}. 
The Galactic spectra typically show either \acet\ or CS 
absorption most prominently.  Figure \ref{fig.ew7} plots the 
partial equivalent width of the absorption band (EW$_{p}$) 
against the apparent centroid of that part of the band
visible in the FORCAST spectra ($\lambda_{p}$).  The figure
also includes the SWS sample, with the spectra analyzed with the
same wavelength stops as the FORCAST spectra.  The majority 
of the CS sources are in the lower left quadrant, with 
$\lambda_{p}$ $<$ 7.25 \mum\ and EW$_{p}$ $<$ 0.005.  The 
\acet\ sources, in contrast, have $\lambda_{p}$ $>$ 7.2 \mum\ 
and EW$_{p}$ $>$ 0, and generally lie in the upper right 
quadrant.  As noted before by \citet{zijlstraea06} and 
\citet{matsuuraea06}, the Magellanic samples do not contain 
any sources with clear CS features.  The vast majority would 
be in the upper right.

The right-hand panel of Figure \ref{fig.ew7} shows the same 
data but color-coded by the variability type.  Almost all of 
the Miras are in the upper right region due to strong \acet\
bands.  The SRbs are split between the CS and \acet\ regions.  
Table \ref{tab.abs} summarizes the distribution of the bands 
between the Miras and SRbs for the Galactic sources.

\begin{deluxetable}{lrrr}
\tablecaption{CS, \acet\ Absorption - Galactic Sample\label{tab.abs}}
\tablewidth{0pt}
\tablehead{
\colhead{} &\multicolumn{3}{c}{No. of Stars showing}\\
\colhead{} & \colhead{\acet} & \colhead{CS}  & \colhead{Other} 
}
\startdata
SRb &  6 (26\%) & 12 (52\%) & 5 (22\%)\\
Mira & 27 (64\%) & 1 (2\%) & 14 (33\%) 
\enddata
\end{deluxetable}

\section{Discussion\label{sec.disc}}

\subsection{Molecular Gas} 

The Galactic carbon stars observed with the SWS and the 
FORCAST grisms show a distinct difference between Miras and
semi-regular variables in their molecular gas features.  
While \acet\ dominates the molecular absorption in the Miras, CS
dominates in about half of the SRb variables.

The Magellanic sample of carbon stars observed with the IRS
do not show this dependence of molecular chemistry on
variability type.  Instead, \acet\ dominates the molecular
absorption, with no CS bands apparent, no matter the
variability type, as first noticed in the LMC by
\citet{zijlstraea06} and \citet{matsuuraea06}.

\citet{matsuuraea06} suggested that the \acet\ band could
mask the CS band in the Magellanic sample, but the bands
occur at different wavelengths and the CS should be 
detectable if it were present.  In addition, we would expect 
a continuous distribution of relative strengths in the two 
bands in the Galactic sample, but Figure \ref{fig.ew7} shows 
a bimodal distribution.

The lack of visible CS bands in the Magellanic spectra likely 
arises from differences in metallicity and lower abundances 
of sulfur.  Metallicity differences can also explain the 
rising strength of the \acet\ bands from the Milky Way to the LMC to the
SMC \citep[e.g.,][]{vlea99,matea02,smcc06}.  The progressively lower oxygen 
abundances in the LMC 
and SMC lead to higher C/O ratios after dredge ups. This in 
turn produces stronger \acet\ bands, as noted previously for 
bands in the 5--8 \mum\ region and $\sim$14 \mum\ 
\citep[e.g.,][]{zijlstraea06,smcc06}, and in the 3--4 \mum\ region 
\citep[e.g.,][]{vlea99, matea02,matsuuraea05,vlea06}.

The addition of the FORCAST spectra to the Galactic carbon star sample 
reveals that with a single exception\footnote{U Cyg, a Mira in Group 2, is
the lone exception.}, CS only appears in the Galactic SRbs, and then only in 
some of them (Table \ref{tab.abs}).  Thus, the presence
of CS requires that the carbon star be (1) metal-rich and 
(2) pulsating as an SRb variable.  However, only half of the semi-regular 
variables in the Galactic sample show CS bands, so these conditions
are necessary, but not sufficient.

\subsection{Dust and Metallicity\label{sec.dustZ}}

The differences in the behavior of the SiC dust emission as a 
function of \smn\ color between the Milky Way and Magellanic 
Clouds are readily apparent in Figure \ref{fig.69silc_mc2}.
Previous works, starting with \cite{smcc06} and 
\cite{zijlstraea06}, largely investigated these differences through 
the lens of metallicity.  The metallicities in the Magellanic Clouds are
typically in the range  [Fe/H] $\sim$ $-$0.7 to $-$0.3 for the LMC 
\citep[e.g.,][]{piattigeisler13} and [Fe/H] $\sim$ $-$1.5 to $-$0.5
in the SMC \citep[e.g.,][]{piatti12, rubeleea15}, and fall even lower for dwarf
spheroidals in the Local Group \citep[e.g.,][ and references therein]{sml12}.
Differences in Si abundances could explain why the relative contribution 
of SiC dust builds only slowly as the dust increases in the 
 SMC and other metal-poor galaxies in the Local Group, while in the 
Galaxy, it rises much more rapidly before the continued formation of
amorphous carbon overwhelms it.

\subsection{Dust and Variability Type\label{sec.dustvar}}

\begin{figure}
\epsscale{1.1}
\includegraphics[width=3.25in]{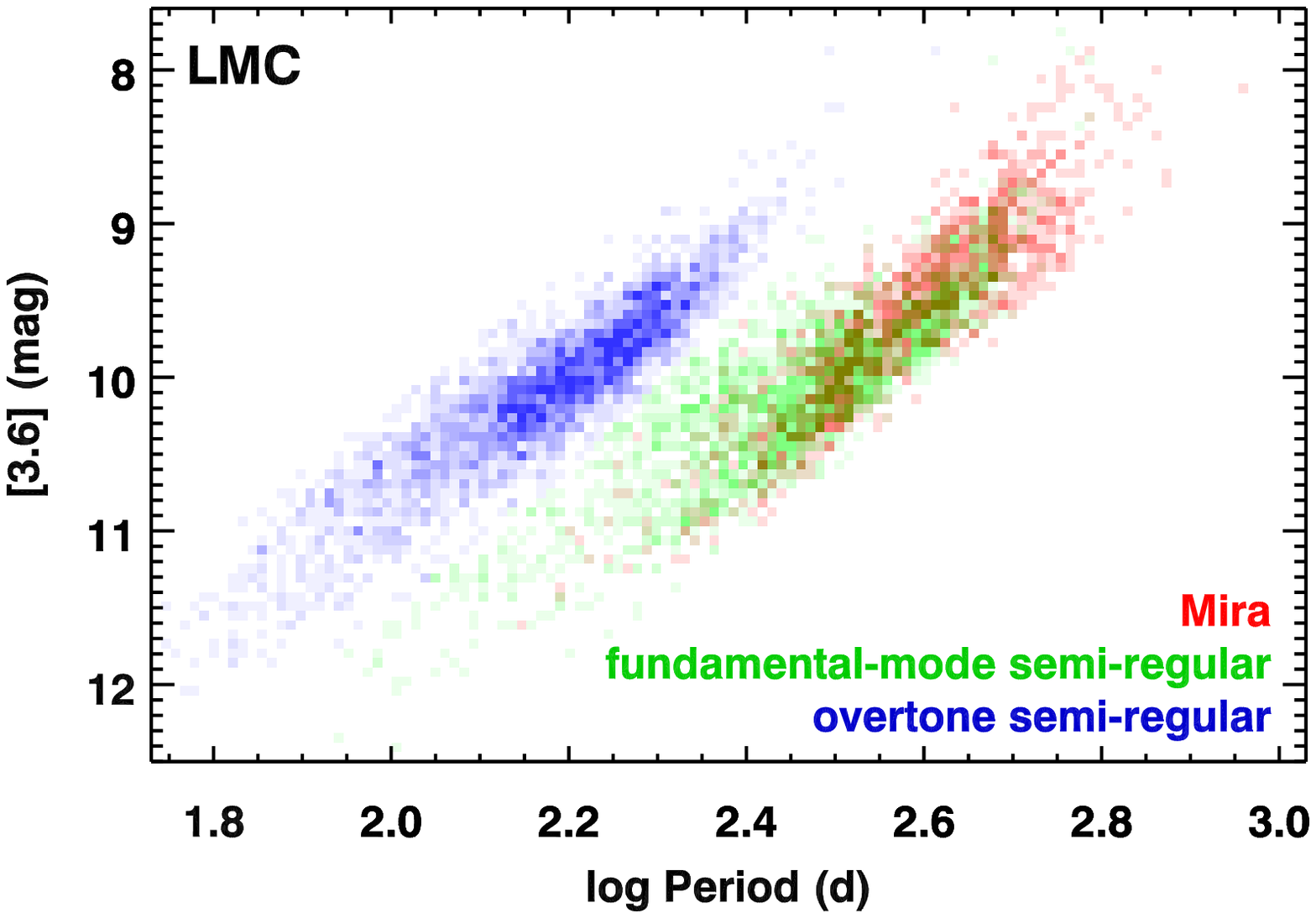}
\includegraphics[width=3.25in]{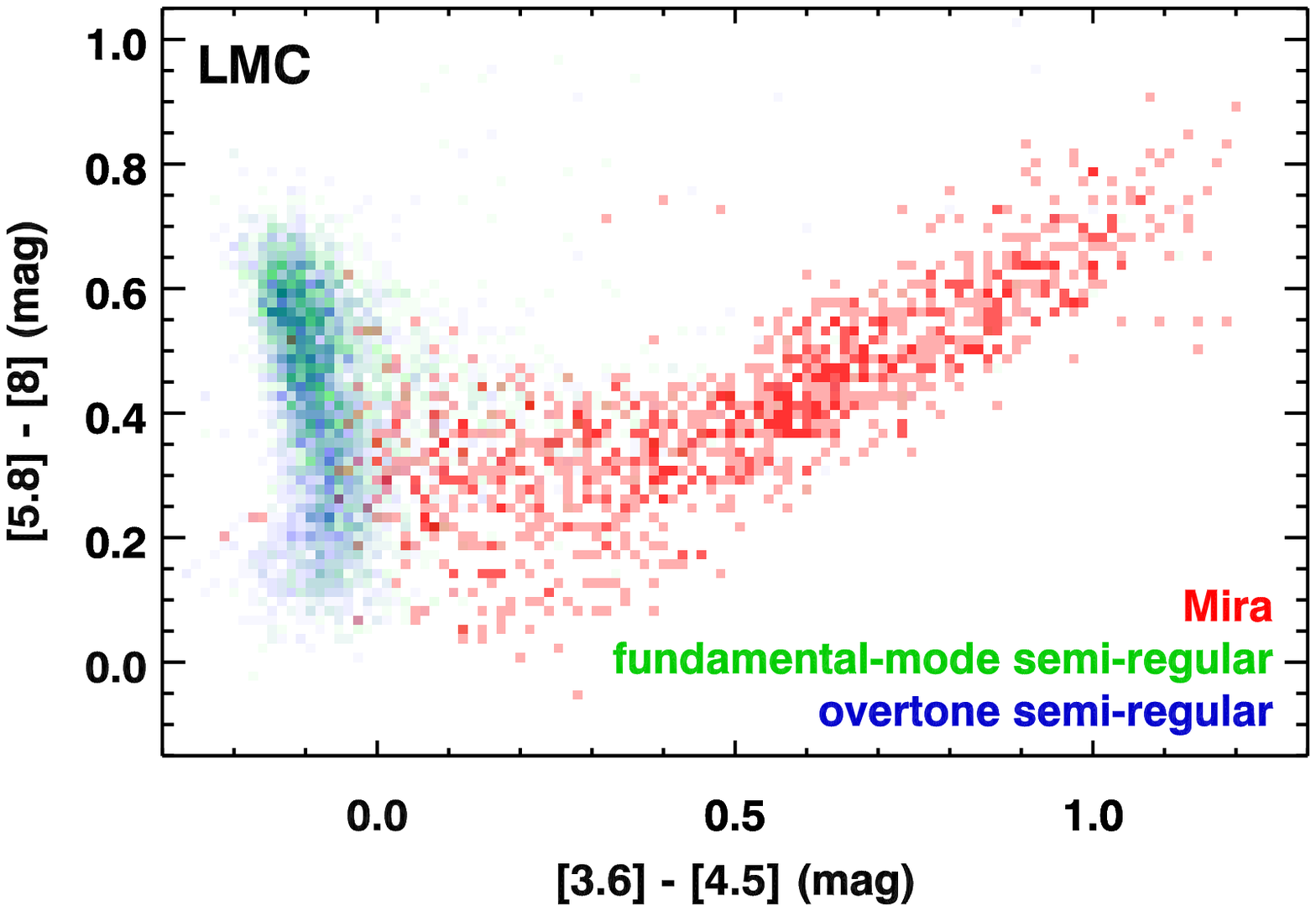}
\caption{LMC carbon stars. (top) Period-luminosity diagram for carbon stars
identified as semi-regular variables on the first overtone sequence (blue),
semi-regulars on the fundamental sequence (green), and Miras (red). (bottom)
Mid-infrared color-color diagram using the same color codes as the P-L diagram.
}
\label{fig.hess}
\end{figure}

With the expanded Galactic sample, we can now revisit the 
dust properties in terms of the variability types.  As
already noted, a sharp color boundary separates the Galactic 
semi-regulars and Miras in Figure \ref{fig.69silc_mw}.
The Magellanic samples observed with the IRS show a similar
dichotomy, with only a handful of exceptions (Fig. \ref{fig.69silc_mc2}). %

Figure \ref{fig.hess} shows related behavior for a much larger 
photometric sample of carbon stars in the LMC from the
OGLE-III survey \citep{sus09}.  The top panel presents a period-luminosity
diagram (P-L)\footnote{Pioneered by \citet{leavitt+1912}.} for  the stars
identified as carbon-rich Miras or semi-regular variables,
with the semi-regulars color-coded according to
whether they appear on the first overtone or the 
fundamental-mode sequence \citep[e.g.,][]{ws96, woodea99, fraserea05}. 
The bottom panel uses
the same color codes and maps the stars into color-color 
space using photometry from the {\it Spitzer} SAGE project
\citep{sage06} and the \wise\ mission \citep{wise10}\footnote{SAGE: Surveying 
Agents of
Galaxy Evolution; \wise: {\em Wide-field Infrared Survey Explorer}.}. The 
 carbon stars in the SMC behave nearly identically \citep{slk15}.

The color-color diagram reveals a clear dichotomy between the 
semi-regulars and Miras.  The carbon stars identified in the 
OGLE-III survey as semi-regulars can pulsate in either an 
overtone or fundamental mode \citep{sus09, swu13}, and thus appear in two 
separate sequences in the P-L diagram. However, nearly all of them 
follow the same sequence in color-color space.  The Mira 
variables follow a different sequence.  
Since amorphous carbon dominates the dust around carbon stars,
adding more dust to a carbon star 
will redden it monotonically in all broadband near-infrared 
and mid-infrared colors, leading to a readily recognizable 
sequence in any color-color space \citep[e.g.,][Sloan et al. in 
press]{sloan17}.  

The [3.6]$-$[4.5] colors 
of the semi-regulars generally stay below 0, although they 
show a range of [5.8]$-$[8] colors.  The relatively blue 
colors of the semi-regulars in all other infrared colors 
indicate that there is too little dust to explain their 
behavior in [5.8]$-$[8].  \citet{slk15} argued that 
the C$_3$ absorption band at $\sim$5 \mum\ is responsible 
for the observed behavior;  other molecules,
such as CO, should also affect the 5.8 \mum\ filter.  In the 
dustier stars, the effect of any molecular absorption bands 
in this spectral region will diminish due to dust veiling.

The additional Galactic carbon stars observed with SOFIA
have revealed a second dependence of dust characteristics on 
the variability type.  The rapid rise of the relative 
strength of the SiC emission feature with increasing overall 
dust content is limited to the semi-regulars.  Once a star begins 
pulsating in the fundamental mode as a Mira, the amorphous 
dust content climbs, pushing the \smn\ color beyond the 
boundary at $\sim$0.3. The relative strength of 
the SiC dust, however, turns over and diminishes. 

\subsection{Variability Type and Pulsation Mode\label{sec.mode}}

\begin{figure}
\includegraphics[width=3.35in]{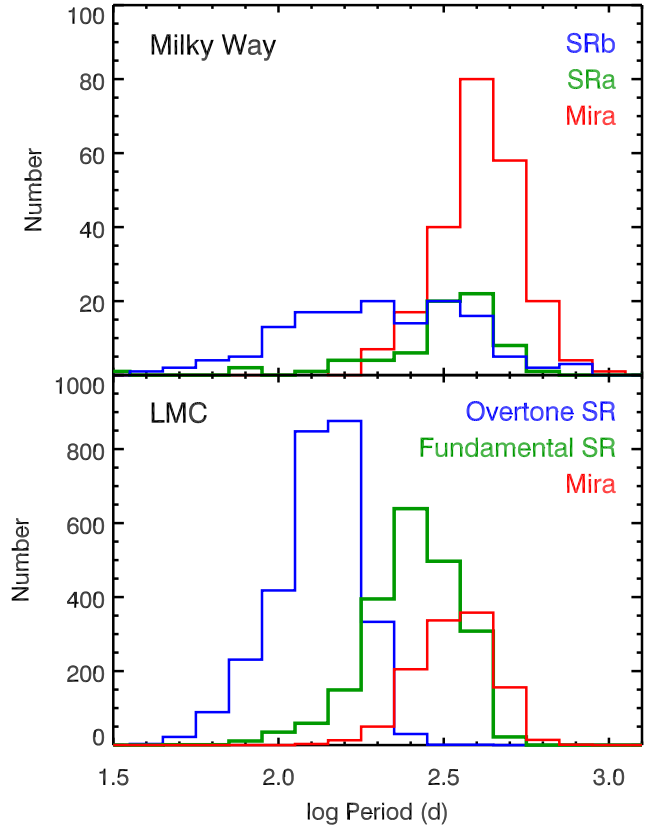}
\caption{Period distributions. (top) Galactic carbon stars classed as SRbs 
(blue), SRas (green), and Miras (red) in the GCVS. (bottom) LMC carbon stars
classed as SRVs on the first overtone sequence (green), SRVs on the fundamental
mode sequence (green), and Miras (red).
}
\label{fig.phist}
\end{figure}

Comparison of the Galactic and Magellanic samples is hampered
by differences in how their variability is classified.  
Pulsating giants in the Galaxy have been classified for 
several decades as Miras if their pulsation amplitude (peak 
to peak) exceeds 2.5 magnitudes at 
$V$ \citep[e.g.,][]{matteiea97,gcvs17}\footnote{\cite{hrw84} 
noted that some observers use a limit of 2 magnitudes, and 
\cite{pgg38} quoted 1.5 magnitudes.}.  The semi-regular 
variables on the AGB have lower amplitudes.  Those with 
fairly stable lightcurves are classified as SRa
variables, while the SRbs are distinguished by more complex
lightcurves that still show a discernible periodicity some
of the time \citep[e.g.,][]{gla68,hrw84}.  

The OGLE-III survey separated Miras from semi-regulars in the 
LMC at an amplitude of $\Delta I = $ 0.8 magnitudes \citep{sus09}.  
Miras pulsate in the
fundamental mode \citep{ws96}. 
The OGLE-III survey did not subdivide the semi-regulars, though, unlike 
many studies of Galactic variables. Figure \ref{fig.hess}
shows that the dominant pulsation mode in semi-regulars can be
either the fundamental mode or first overtone, as has been noted
previously with shorter-wavelength data \citep[e.g.,][]{woodea99}.  

Thus, we have two groups of carbon-rich semi-regular variables 
in the LMC, and two in the Galaxy.  The Galactic SRas have 
lightcurves with well-defined periodicities but weak 
amplitudes, making them the Galactic analogs of
the fundamental-mode semi-regulars in the LMC.  Similarly, 
Galactic SRbs are  overtone pulsators.  Using Gaia distances for 
carbon stars has proven challenging so far \cite[e.g.,][]{mcdonaldea18}, so
placing the SRa and SRb variables on a P-L diagram is problematic.

We can still compare the periods, though.  Figure \ref{fig.phist} shows
that the distribution of periods for the SRb, SRa, and Mira 
variables  in the Galaxy behave very much like the
overtone semi-regulars, fundamental-mode 
semi-regulars, and Miras in the LMC, respectively.  For the Galactic 
sample, we cross-referenced the Catalog of Galactic Carbon 
Stars \citep[3rd Ed.;][]{cgcs01} with the General Catalog of 
Variable Stars \citep[GCVS 5.1;][]{gcvs17}, matching sources 
within 10$\arcsec$.  The quantitative differences in the
distributions likely arise from the differences in the 
initial mass limits for carbon stars in the LMC and Galaxy,
which in turn limit the range of luminosities a star will
show on the AGB\footnote{That is, the lowest mass carbon stars in the
LMC remain oxygen-rich in the Galaxy.}. The P-L relation ties these 
luminosities to the pulsation periods of the stars.

The qualitative conclusion remains clear:  the first overtone
dominates the pulsation of the SRb variables, while the 
fundamental mode has grown stronger in the SRa variables.
Once the fundamental mode dominates all other modes and the
amplitude grows strong enough, the star becomes a Mira
variable.

\subsection{Pulsation and Dust}

We can now relate the differences in dust quantity and
composition illustrated in Figures \ref{fig.69silc_mw}
and \ref{fig.69silc_mc2} with the
pulsation mode of the central star.  Stars with weak
pulsations do not produce significant quantities of dust.
Whether their pulsations are dominated by an overtone mode 
or the  fundamental, if their amplitudes are less than 
$\sim$2.5 mag.\ in $V$, their \smn\ colors remain below 0.3, 
always in the Galaxy and most of the time in the Magellanic
Clouds.  For most of the carbon-rich semi-regulars in the
LMC and all in the Galaxy, a slight increase in total dust
content is accompanied by a sharp increase in the amount of
SiC dust.  Once the fundamental pulsation mode dominates
the star {\it and} its pulsation amplitude grows strong
enough for it to be classified as a Mira, the total dust
content takes off, and the relative strength of the SiC
feature diminishes.

To rephrase, the key to significant dust production on the 
AGB is not just the pulsation mode, but also the pulsation 
amplitude of the central star \cite[e.g.,][]{mattssonea08,
mcdtrabucchi19}. 
These strong pulsations are sufficient to push molecular gas 
far enough from the stellar photosphere so that amorphous 
carbon can condense in the quantities necessary for radiation 
pressure on the dust to drive significant mass loss
\cite[e.g.,][]{mcdonaldea18, mcdtrabucchi19}.
 The forming dust will incorporate much of the 
carbon-rich molecular gas in the outflows and veil the
absorption from those that are left, which will weaken the
molecular absorption as the dust content grows.

When overtone modes dominate the fundamental mode, the
pulsation amplitude remains weak, and the star is unable 
to drive significant dust production or mass loss.  In
these cases, the \smn\ color is limited by the lack of
overall dust, and the SiC dust emission feature is relatively
strong.

The semi-regular variables on the fundamental-mode sequence may be 
of too low an initial mass to drive strong pulsations. Alternatively,
they may have transitioned to this 
sequence too recently and have not yet had time for their pulsations to 
grow enough to trigger dust production. Thus, their dust 
properties often still look like the semi-regulars on the 
first overtone (SRbs) even though their pulsation periods and 
luminosities place them on the fundamental sequence with the 
Miras in the P-L diagrams.

\subsection{Grain Scenarios}

\cite{lag07} suggested that the two sequences of SiC versus 
total dust (Fig. \ref{fig.69silc_mc2}) differ in part due to the 
metallicity dependence
of the condensation temperatures of SiC and graphite.  
Building on this idea, \cite{leisenringea08} proposed that 
carbon-rich dust grains could form in layers 
\citep[e.g.,][]{koz96, lor01}, and the upper sequence resulted 
from the initial condensation of SiC dust which forms first 
 \cite[e.g.,][]{mccabe82,fg02,dellagliea17, nanniea19}.
These grains would then 
be covered by a layer of amorphous carbon, which progressively hides 
the SiC from the observer.

Another possibility should be considered.  The SiC and
amorphous carbon could form in separate populations of 
grains.  In that case, the diminishing contribution of the
SiC dust once stars start pulsating as Miras can be explained
simply by the decreasing fraction of SiC dust as the 
condensation of amorphous carbon takes off.  The relative
abundances of carbon and silicon should make that result 
inevitable even without layering. As carbon is synthesized in the
thermal pulses and silicon is not, the relative abundances become
even more skewed as the star evolves and dredges up freshly fused carbon.

The infrared spectra provide no means of distinguishing 
whether the SiC and amorphous carbon grains are layered or
form separately.  In either scenario, the sharp boundary 
between the Miras and the semi-regulars at [6]$-$[9] $\sim$ 
0.3 and the dramatic difference in the behavior of the 
relative strength of the SiC feature on either side of 
that boundary point to a clear difference in the dust-formation mechanism 
between the Miras and semi-regulars.

\citet{croatea05} used isotopic abundances and the relative
lack of s-process elements to show that SiC grains in 
meteorites condensed at an earlier stage in AGB evolution 
than other carbonaceous grains and under different 
conditions. Our finding that 
the SiC dust is produced predominantly by weakly pulsating carbon stars 
supports their conclusion.  The different isotopic abundances
in the SiC grains found by Croat et al. support the commonly assumed sequence
of stars evolving from semi-regulars to Miras.

\subsection{The Beginning of the End?}

The end of a star's life on the AGB must occur after it has lost
a substantial fraction of its mass via a stellar wind driven
by pulsation and radiation pressure on the condensing
dust \cite[e.g.,][]{whitelockea03,mcdonaldea18, bladhea19}. The pulsations
in the stellar interior can levitate material into the dust-formation zone,
where the increased molecular and dust opacity allow the radiation to 
drive the wind \citep[as first laid out by ][see also \citet{ho18} and 
references therein]{jonesea81}.
What physical process triggers the pulsations which drive heavy mass 
loss, though, is still a matter of debate. For carbon stars, the C/O
ratio is an important parameter in the wind properties, although a
better description may be that
the total amount of excess free carbon is the key property 
\cite[e.g.,][]{lz08, sml12, erikssonea14, nanniea17, bladhea19}. 
That is, the trigger would be a dredge-up event that pushes 
the C/O ratio or the amount of free carbon over a critical threshold.

\citet{mcdtrabucchi19}, though, found that a transition between segments 
of the first overtone sequence, indicated by a larger $I$-band 
amplitude, was tied to significantly increased mass-loss.
Here, the dramatic change we observe in dust and gas properties
between the first-overtone semi-regulars and the fundamental-mode Miras 
 suggest that the transition of a carbon star to a Mira variable also 
triggers a substantial increase in dust production 
and total  mass loss. This increase will rapidly strip its 
envelope and end its life on the AGB \cite[e.g.,][]{mattssonea08,mcdonaldea18}.

{

The relations between the pulsational behavior of carbon stars and the
properties of the gas and dust around them raise questions about
oxygen-rich AGB stars, where the dust is usually dominated by
silicates.  For these stars, an infrared spectroscopic study comparing
the overall dust quantity between the Miras and semi-regulars analogous
to what we have done in this paper has yet to be done.
Differences in mineralogy have been found, though, and
point to significantly different dust processing in the semi-regulars
\cite[e.g.,][]{sloanea96, sloan13um, uttea19}.

Recent radio observations have revealed details on molecular line profiles 
and expansion velocities in the outer shell around both carbon-rich and
oxygen-rich stars \cite[e.g.,][McDonald et al. submitted]{mcdonaldea18, 
dlea19, massalkhiea19}.  \citet{dlea19}, for example, have recently found 
that some oxygen-rich semi-regulars have an unusual CO line profile 
inconsistent with spherical symmetry.  This result needs to be confirmed 
in a larger sample, especially to confirm that it only appears in the 
oxygen-rich semi-regulars.
Additional observations and archival studies in both the infrared
and radio are needed to better understand the relation between the
pulsation of the central star, its chemistry, and how it ejects its
envelope, forms dust, and evolves off of the AGB.
}

\section{Summary \& Conclusions \label{sec.summ}}

We observed 33 Galactic carbon stars from 5 to 13.8 \mum\ with SOFIA's
FORCAST grisms. The sources were selected to expand the Galactic sample
observed with the SWS on \iso\ and better align it with the Magellanic
samples observed by the IRS on \spi. Using the Manchester system,
we extracted the strengths of the dust features
and equivalent widths for molecular gas
absorption bands, with the wavelength ranges adjusted to avoid telluric
features.

The new Galactic sample reveals multiple differences in the spectral
properties between carbon-rich semi-regular and Mira variables.
 The semi-regulars in the sample include 
both SRb and SRa variables, in which the pulsations are dominated by 
the first overtone and fundamental mode, respectively. The Miras, also
fundamental-mode pulsators, have much stronger pulsations than the SRas.

A [6]$-$[9] color of $\sim$0.3 cleanly separates the Miras from the 
semi-regulars. That color tracks the amount of amorphous carbon dust 
around the star and
shows that the semi-regulars have little circumstellar dust
compared to the Miras.

The strength of the SiC dust feature at $\sim$11.3 \mum\
rises sharply in the semi-regular variables.  Once a star
becomes a Mira, it begins to produce amorphous carbon dust in
substantial quantities, and the SiC dust feature is masked
as the overall dust content grows.

All of these differences between the semi-regulars and Miras
support the argument that strong fundamental-mode pulsations
are required for significant rates of dust-production in
carbon stars.  This statement holds for both the Galactic
sample studied here and the Magellanic samples as well.

The 5--7.5 \mum\ portion of the spectra can show absorption bands
from CS and C$_2$H$_2$, but the CS is nearly always absent among the
Miras, while it is present in half of the semi-regulars.
That is, CS only occurs in stars with weak
pulsations, and then only in some of them.  
CS is not observed
in Magellanic carbon stars regardless of their variability type;
 its presence must also require higher metallicity.

Lastly, for the Galactic sample, the clean separation of the Miras
and semi-regulars in [6]$-$[9] color points to a simple
means of distinguishing these types of carbon-rich variables.  
Single-epoch infrared photometry or spectra can quantify the amount of 
dust around these stars and that makes it possible to identify the Miras
without measuring their pulsation amplitudes or periods, both
of which require observations of long temporal baselines, or
placing them on a period-luminosity diagram, which requires
knowing their distances.

\acknowledgments

{ We thank the anonymous referee whose careful reading and thoughtful 
suggestions have helped improve the paper. We also thank the SOFIA flight crew
and staff scientists for making these observations.}
Based on observations made with the NASA/DLR Stratospheric 
Observatory for Infrared Astronomy (SOFIA). SOFIA is jointly operated by the 
Universities Space Research Association, Inc. (USRA), under NASA contract 
NNA17BF53C, and the Deutsches SOFIA Institut (DSI) under DLR contract 
50 OK 0901 to the University of Stuttgart. Financial support for this work 
was provided by NASA through awards SOF03-0104 and 
SOF04-0129 issued by USRA.
We made use of the
NASA Astrophysics Data System, IRSA's Gator service, and CDS's Simbad and
Vizier services. 

\facilities{SOFIA (FORCAST), \spi~(IRS), \iso~(SWS)}

\clearpage

\appendix
\section{Adjustments to Feature Extraction Wavelengths and Their 
Effects\label{app.lambdas}}

The Manchester method works as follows. The continuum is determined at
feature-free wavelengths by averaging the flux in small wavelength ranges.
As noted in the main text, residuals from atmospheric ozone precludes the
use of the same wavelength ranges that \cite{mcc16} used for their feature
extraction for two of the four wavelengths. Specifically, the 9 \mum\
end of the [6]$-$[9] color has to be slightly bluer,  and the short-wavelength
end of the
SiC feature has to be somewhat redder than the original set.   

To determine the new locations, the ozone in the SOFIA data needs to be 
avoided, along with the SiC feature. We also ensure that 
the IRS data have enough data points to be
valid, since those spectra are not as over-sampled as the SOFIA data are. 
The SWS data are at higher spectral resolution, and thus this is not an 
issue for them (the smallest number of data points in a given wavelength range
is 71). 
This changes the
color from [6.4]$-$[9.3] to [6.4]$-$[9.1]. The blue end of the SiC feature 
changes from
$\sim$9.8 to $\sim$10.18 \mum. The [6.4] band and the red end of the SiC 
feature remain unchanged. Table \ref{tab.lambdas} gives the original and new
wavelengths ranges.

\begin{figure}
\includegraphics[width=3.35in]{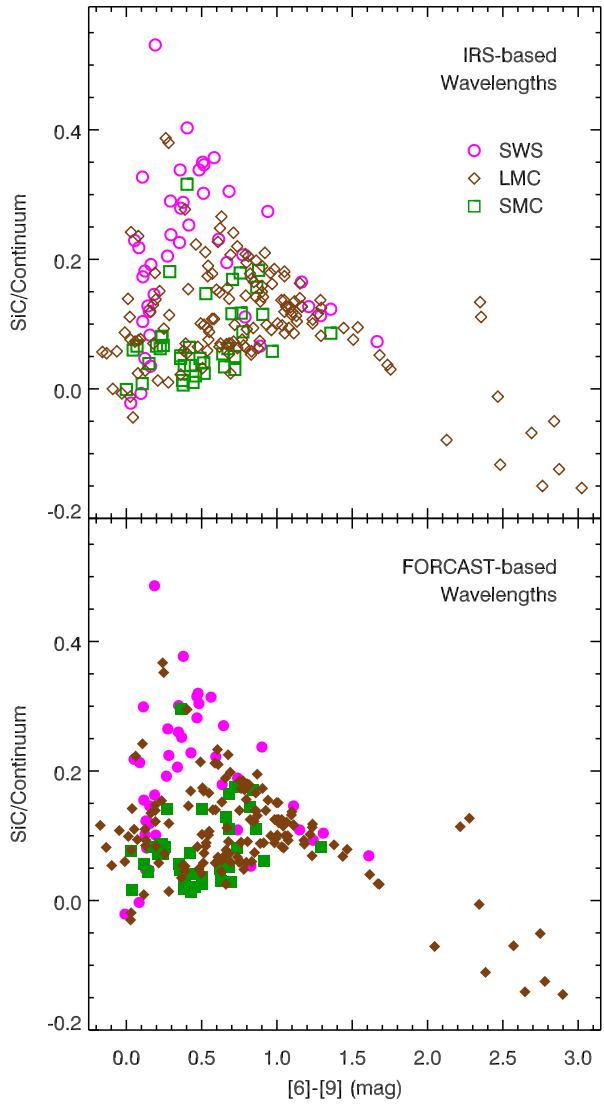}
\caption{Comparisons of feature extractions using the FORCAST (FC) and IRS
wavelength ranges. (top) [6]$-$[9] vs. SiC/Cont. for the original, IRS-based 
ranges. (bottom) [6]$-$[9] vs. SiC/Cont. for the new, FORCAST-based 
ranges. The magenta circles are the SWS sample; the brown diamonds are the 
IRS sample from the LMC; and the green squares are the IRS sample from the
SMC.
}
\label{fig.sic69comp1}
\end{figure}

The features strengths for the SWS and IRS samples based on the new wavelength 
ranges are given in Tables \ref{tab.new69sws} and \ref{tab.new69irs}, 
respectively. The values from the previous
extractions \citep{mcc16} are also given for comparison.  Figure \ref{fig.sic69comp1} shows
the [6]$-$[9] vs. SiC/Continuum for the original ranges (top panel) and for 
the new ranges (bottom panel). It
is evident that the two plots are qualitatively the same, even though some 
data points have shifted slightly. We conclude that the new ranges provide 
reliable results for the analysis in the main part of this work.

\begin{figure}
\includegraphics[width=3.35in]{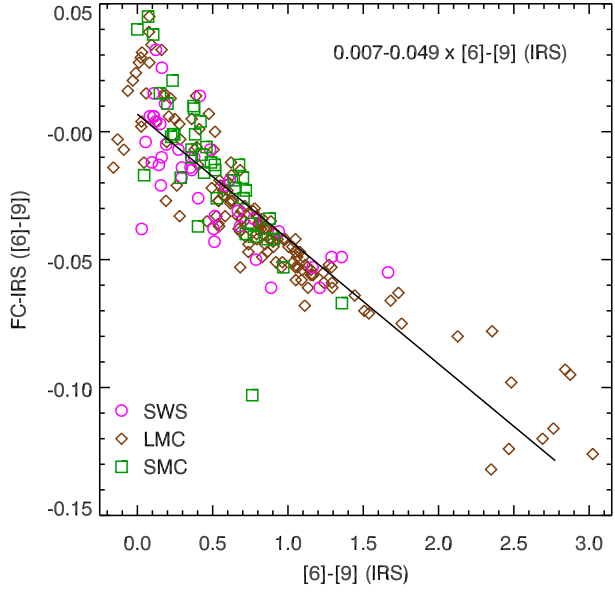}
\caption{Difference between the [6]$-$[9] color from the FORCAST wavelength
ranges and that from the IRS ranges as a function of the IRS-derived color.
Symbols are the same as Fig. \ref{fig.sic69comp1}. The solid line is a
linear fit to the data, and its parameters are given in the upper right
as well as Table \ref{tab.lin69}.}
\label{fig.sic69comp2}
\end{figure}

There are a few trends that should be noted. The 
[6]$-$[9] colors from the new range are slightly bluer than the original 
colors, and the magnitude of the difference is a function of the color. 
The strength of the SiC/Continuum is slightly smaller with the new
ranges, although this difference is not nearly as tight a function of the
strength as the color difference is. Figure \ref{fig.sic69comp2} shows
the differences as functions of the original values.

We fit a line to the difference in color as a function of color for
(a) the SWS data, (b) the LMC IRS data, 
(c) the SMC IRS data, and (d) the full SWS + IRS dataset, the results of 
which are given in Table \ref{tab.lin69}.
These can be used to adjust the results for other datasets whose features
were extracted using the original wavelength ranges.

\clearpage
\begin{deluxetable}{lcc}
\tablecaption{Extraction Range Comparison\label{tab.lambdas}}
\tablewidth{0pt}
\tablehead{
\colhead{} &\multicolumn{2}{c}{$\lambda$ Range}\\
\colhead{} &\colhead{Blue} & \colhead{Red}
}
\startdata
\multicolumn{3}{c}{[6]$-$[9]} \\\hline
Original & 6.25--6.55 \micron& 9.10--9.50 \micron\\
New      & 6.25--6.55 \micron& 9.00--9.21 \micron\\
\hline
\multicolumn{3}{c}{SiC/Continuum}\\\hline
Original & 9.50--10.10 \micron& 12.50--12.90 \micron\\
New      & 10.11--10.25 \micron& 12.50--12.90 \micron\\
\enddata
\end{deluxetable}

\begin{deluxetable}{lRRRR}
\tablecaption{Revised Feature Strengths - SWS Sample\label{tab.new69sws}}
\tablewidth{0pt}
\tablehead{
\colhead{} & \multicolumn{2}{c}{[6]$-$[9] (mag) } & 
\multicolumn{2}{c}{SiC/Continuum}\\
\colhead{Source} & \colhead{Revised} & \colhead{Original} & 
\colhead{Revised} & \colhead{Original}
}
\startdata
WZ Cas &-0.009 \pm 0.002 &  0.029 \pm 0.002 &  -0.021 \pm 0.003 & -0.022 \pm 0.002  \\
VX And & 0.155 \pm 0.003 &  0.123 \pm 0.003 &   0.146 \pm 0.003 &  0.182 \pm 0.001  \\
HV Cas & 0.339 \pm 0.002 &  0.353 \pm 0.002 &   0.206 \pm 0.001 &  0.226 \pm 0.001  \\
R Scl  & 0.266 \pm 0.002 &  0.273 \pm 0.002 &   0.192 \pm 0.001 &  0.205 \pm 0.001
\enddata
\tablecomments{This table is available in its entirety in machine-readable 
format. A portion is shown here for guidance regarding its form and content.}
\end{deluxetable}

\begin{deluxetable}{lCCCC}
\tablecaption{Revised Feature Strengths - IRS Sample\label{tab.new69irs}}
\tablewidth{0pt}
\tablehead{
\colhead{} & \multicolumn{2}{c}{[6]$-$[9]} & \multicolumn{2}{c}{SiC/Cont.}\\
\colhead{Source} & \colhead{Revised} & \colhead{Original} & 
\colhead{Revised} & \colhead{Original}
}
\startdata
MSX SMC 33 & 0.501 \pm 0.007 & 0.514 \pm 0.009 & 0.036 \pm 0.002 & 0.040 \pm 0.002 \\
MSX SMC 36 & 0.733 \pm 0.005 & 0.774 \pm 0.005 & 0.081 \pm 0.003 & 0.088 \pm 0.004 \\
MSX SMC 44 & 0.502 \pm 0.011 & 0.517 \pm 0.006 & 0.025 \pm 0.005 & 0.024 \pm 0.004 \\
MSX SMC 54 & 0.720 \pm 0.002 & 0.756 \pm 0.004 & 0.175 \pm 0.004 & 0.179 \pm 0.003 \\
\enddata
\tablecomments{This table is available in its entirety in machine-readable 
format. A portion is shown here for guidance regarding its form and content.}
\end{deluxetable}

\begin{deluxetable}{crr}
\tablecaption{Linear Fits\label{tab.lin69}}
\tablewidth{0pt}
\tablehead{
\colhead{Sample} &\colhead{FORCAST $-$ IRS [6]$-$[9]} & \colhead{N$_{stars}$}
}
\startdata
SWS & $\Delta$ = 0.003 $-~0.046~\times$ (IRS [6]$-$[9]) & 42\\
LMC & $\Delta$ = 0.006 $-~0.047~\times$ (IRS [6]$-$[9]) & 144\\
SMC & $\Delta$ = 0.025 $-~0.078~\times$ (IRS [6]$-$[9]) & 40\\
All & $\Delta$ = 0.007 $-~0.049~\times$ (IRS [6]$-$[9]) & 226
\enddata
\end{deluxetable}

\bibliographystyle{aasjournal}
\bibliography{fccrefs}

\begin{thebibliography}{}
\expandafter\ifx\csname natexlab\endcsname\relax\def\natexlab#1{#1}\fi
\providecommand{\url}[1]{\href{#1}{#1}}
\providecommand{\dodoi}[1]{doi:~\href{http://doi.org/#1}{\nolinkurl{#1}}}
\providecommand{\doeprint}[1]{\href{http://ascl.net/#1}{\nolinkurl{http://ascl%
.net/#1}}}
\providecommand{\doarXiv}[1]{\href{https://arxiv.org/abs/#1}{\nolinkurl{https:%
//arxiv.org/abs/#1}}}

\bibitem[{{Alksnis} {et~al.}(2001){Alksnis}, {Balklavs}, {Dzervitis},
  {Eglitis}, {Paupers}, \& {Pundure}}]{cgcs01}
{Alksnis}, A., {Balklavs}, A., {Dzervitis}, U., {et~al.} 2001, Baltic
  Astronomy, 10, 1, \dodoi{10.1515/astro-2001-1-202}

\bibitem[{{Aoki} {et~al.}(1998){Aoki}, {Tsuji}, \& {Ohnaka}}]{aokiea98}
{Aoki}, W., {Tsuji}, T., \& {Ohnaka}, K. 1998, \aap, 340, 222

\bibitem[{{Aoki} {et~al.}(1999){Aoki}, {Tsuji}, \& {Ohnaka}}]{aokiea99}
---. 1999, \aap, 350, 945

\bibitem[{{Bladh} {et~al.}(2019){Bladh}, {Eriksson}, {Marigo}, {Liljegren}, \&
  {Aringer}}]{bladhea19}
{Bladh}, S., {Eriksson}, K., {Marigo}, P., {Liljegren}, S., \& {Aringer}, B.
  2019, \aap, 623, A119, \dodoi{10.1051/0004-6361/201834778}

\bibitem[{{Boyer} {et~al.}(2011){Boyer}, {Srinivasan}, {van Loon}, {McDonald},
  {Meixner}, {Zaritsky}, {Gordon}, {Kemper}, {Babler}, {Block}, {Bracker},
  {Engelbracht}, {Hora}, {Indebetouw}, {Meade}, {Misselt}, {Robitaille},
  {Sewi{\l}o}, {Shiao}, \& {Whitney}}]{boyerea11}
{Boyer}, M.~L., {Srinivasan}, S., {van Loon}, J.~T., {et~al.} 2011, \aj, 142,
  103, \dodoi{10.1088/0004-6256/142/4/103}

\bibitem[{{Croat} {et~al.}(2005){Croat}, {Stadermann}, \&
  {Bernatowicz}}]{croatea05}
{Croat}, T.~K., {Stadermann}, F.~J., \& {Bernatowicz}, T.~J. 2005, The
  Astrophysical Journal, 631, 976, \dodoi{10.1086/432598}

\bibitem[{{de Graauw} {et~al.}(1996){de Graauw}, {Haser}, {Beintema},
  {Roelfsema}, {van Agthoven}, {Barl}, {Bauer}, {Bekenkamp}, {Boonstra},
  {Boxhoorn}, {Cote}, {de Groene}, {van Dijkhuizen}, {Drapatz}, {Evers},
  {Feuchtgruber}, {Frericks}, {Genzel}, {Haerendel}, {Heras}, {van der Hucht},
  {van der Hulst}, {Huygen}, {Jacobs}, {Jakob}, {Kamperman}, {Katterloher},
  {Kester}, {Kunze}, {Kussendrager}, {Lahuis}, {Lamers}, {Leech}, {van der
  Lei}, {van der Linden}, {Luinge}, {Lutz}, {Melzner}, {Morris}, {van Nguyen},
  {Ploeger}, {Price}, {Salama}, {Schaeidt}, {Sijm}, {Smoorenburg}, {Spakman},
  {Spoon}, {Steinmayer}, {Stoecker}, {Valentijn}, {Vandenbussche}, {Visser},
  {Waelkens}, {Waters}, {Wensink}, {Wesselius}, {Wiezorrek}, {Wieprecht},
  {Wijnbergen}, {Wildeman}, \& {Young}}]{sws96}
{de Graauw}, T., {Haser}, L.~N., {Beintema}, D.~A., {et~al.} 1996, \aap, 315,
  L49

\bibitem[{{Dell'Agli} {et~al.}(2017){Dell'Agli}, {Garc{\'\i}a-Hern{\'a}ndez},
  {Schneider}, {Ventura}, {La Franca}, {Valiante}, {Marini}, \& {Di
  Criscienzo}}]{dellagliea17}
{Dell'Agli}, F., {Garc{\'\i}a-Hern{\'a}ndez}, D.~A., {Schneider}, R., {et~al.}
  2017, Monthly Notices of the Royal Astronomical Society, 467, 4431,
  \dodoi{10.1093/mnras/stx387}

\bibitem[{{Dell'Agli} {et~al.}(2019){Dell'Agli}, {Valiante}, {Kamath},
  {Ventura}, \& {Garc{\'\i}a-Hern{\'a}ndez}}]{dellagliea19}
{Dell'Agli}, F., {Valiante}, R., {Kamath}, D., {Ventura}, P., \&
  {Garc{\'\i}a-Hern{\'a}ndez}, D.~A. 2019, \mnras, 486, 4738,
  \dodoi{10.1093/mnras/stz1164}

\bibitem[{{D{\'\i}az-Luis} {et~al.}(2019){D{\'\i}az-Luis}, {Alcolea},
  {Bujarrabal}, {Santand er-Garc{\'\i}a}, {Castro-Carrizo},
  {G{\'o}mez-Garrido}, \& {Desmurs}}]{dlea19}
{D{\'\i}az-Luis}, J.~J., {Alcolea}, J., {Bujarrabal}, V., {et~al.} 2019, \aap,
  629, A94, \dodoi{10.1051/0004-6361/201936087}

\bibitem[{{Eriksson} {et~al.}(2014){Eriksson}, {Nowotny}, {H{\"o}fner},
  {Aringer}, \& {Wachter}}]{erikssonea14}
{Eriksson}, K., {Nowotny}, W., {H{\"o}fner}, S., {Aringer}, B., \& {Wachter},
  A. 2014, \aap, 566, A95, \dodoi{10.1051/0004-6361/201323241}

\bibitem[{{Ferrarotti} \& {Gail}(2002)}]{fg02}
{Ferrarotti}, A.~S., \& {Gail}, H.~P. 2002, Astronomy and Astrophysics, 382,
  256, \dodoi{10.1051/0004-6361:20011580}

\bibitem[{{Fraser} {et~al.}(2005){Fraser}, {Hawley}, {Cook}, \&
  {Keller}}]{fraserea05}
{Fraser}, O.~J., {Hawley}, S.~L., {Cook}, K.~H., \& {Keller}, S.~C. 2005, \aj,
  129, 768, \dodoi{10.1086/426749}

\bibitem[{{Glasby}(1968)}]{gla68}
{Glasby}, J.~S. 1968, {Variable stars}

\bibitem[{{Goebel} {et~al.}(1981){Goebel}, {Bregman}, {Witteborn}, {Taylor}, \&
  {Willner}}]{goebelea81}
{Goebel}, J.~H., {Bregman}, J.~D., {Witteborn}, F.~C., {Taylor}, B.~J., \&
  {Willner}, S.~P. 1981, \apj, 246, 455, \dodoi{10.1086/158944}

\bibitem[{{Goebel} {et~al.}(1980){Goebel}, {Bregman}, {Goorvitch}, {Strecker},
  {Puetter}, {Russell}, {Soifer}, {Willner}, {Forrest}, \&
  {Houck}}]{goebelea80}
{Goebel}, J.~H., {Bregman}, J.~D., {Goorvitch}, D., {et~al.} 1980, \apj, 235,
  104, \dodoi{10.1086/157615}

\bibitem[{{Groenewegen} \& {Sloan}(2018)}]{gs18}
{Groenewegen}, M.~A.~T., \& {Sloan}, G.~C. 2018, \aap, 609, A114,
  \dodoi{10.1051/0004-6361/201731089}

\bibitem[{{Groenewegen} {et~al.}(2007){Groenewegen}, {Wood}, {Sloan},
  {Blommaert}, {Cioni}, {Feast}, {Hony}, {Matsuura}, {Menzies}, {Olivier},
  {Vanhollebeke}, {van Loon}, {Whitelock}, {Zijlstra}, {Habing}, \&
  {Lagadec}}]{groenewegenea07}
{Groenewegen}, M.~A.~T., {Wood}, P.~R., {Sloan}, G.~C., {et~al.} 2007, \mnras,
  376, 313, \dodoi{10.1111/j.1365-2966.2007.11428.x}

\bibitem[{{Habing}(1996)}]{hab96}
{Habing}, H.~J. 1996, \aapr, 7, 97, \dodoi{10.1007/PL00013287}

\bibitem[{Herter {et~al.}(2012)Herter, Adams, Buizer, Gull, Schoenwald,
  Henderson, Keller, Nikola, Stacey, \& Vacca}]{forcast}
Herter, T.~L., Adams, J.~D., Buizer, J. M.~D., {et~al.} 2012, The Astrophysical
  Journal Letters, 749, L18.
\newblock \url{http://stacks.iop.org/2041-8205/749/i=2/a=L18}

\bibitem[{{Hoffmeister} {et~al.}(1984){Hoffmeister}, {Richter}, \&
  {Wenzel}}]{hrw84}
{Hoffmeister}, C., {Richter}, G., \& {Wenzel}, W. 1984, {Veraenderliche Sterne}

\bibitem[{{H{\"o}fner} \& {Olofsson}(2018)}]{ho18}
{H{\"o}fner}, S., \& {Olofsson}, H. 2018, \aapr, 26, 1,
  \dodoi{10.1007/s00159-017-0106-5}

\bibitem[{{Houck} {et~al.}(2004){Houck}, {Roellig}, {van Cleve}, {Forrest},
  {Herter}, {Lawrence}, {Matthews}, {Reitsema}, {Soifer}, {Watson}, {Weedman},
  {Huisjen}, {Troeltzsch}, {Barry}, {Bernard-Salas}, {Blacken}, {Brandl},
  {Charmandaris}, {Devost}, {Gull}, {Hall}, {Henderson}, {Higdon}, {Pirger},
  {Schoenwald}, {Sloan}, {Uchida}, {Appleton}, {Armus}, {Burgdorf},
  {Fajardo-Acosta}, {Grillmair}, {Ingalls}, {Morris}, \& {Teplitz}}]{irs04}
{Houck}, J.~R., {Roellig}, T.~L., {van Cleve}, J., {et~al.} 2004, \apjs, 154,
  18, \dodoi{10.1086/423134}

\bibitem[{{Jones} {et~al.}(1990){Jones}, {Bryja}, {Gehrz}, {Harrison},
  {Johnson}, {Klebe}, \& {Lawrence}}]{jonesea90}
{Jones}, T.~J., {Bryja}, C.~O., {Gehrz}, R.~D., {et~al.} 1990, \apjs, 74, 785,
  \dodoi{10.1086/191518}

\bibitem[{{Jones} {et~al.}(1981){Jones}, {Ney}, \& {Stein}}]{jonesea81}
{Jones}, T.~W., {Ney}, E.~P., \& {Stein}, W.~A. 1981, \apj, 250, 324,
  \dodoi{10.1086/159378}

\bibitem[{{J{\o}rgensen} {et~al.}(2000){J{\o}rgensen}, {Hron}, \&
  {Loidl}}]{jorgensenea00}
{J{\o}rgensen}, U.~G., {Hron}, J., \& {Loidl}, R. 2000, \aap, 356, 253

\bibitem[{{Jura}(1983)}]{jura83a}
{Jura}, M. 1983, \apj, 267, 647, \dodoi{10.1086/160901}

\bibitem[{{Jura}(1986)}]{jura86}
---. 1986, \apj, 303, 327, \dodoi{10.1086/164077}

\bibitem[{{Keller} {et~al.}(2010){Keller}, {Deen}, {Jaffe}, {Ennico}, {Greene},
  {Adams}, {Herter}, \& {Sloan}}]{kellerea10}
{Keller}, L., {Deen}, C.~P., {Jaffe}, D.~T., {et~al.} 2010, in Society of
  Photo-Optical Instrumentation Engineers (SPIE) Conference Series, Vol. 7735,
  \procspie, 77356N, \dodoi{10.1117/12.857127}

\bibitem[{{Kessler} {et~al.}(1996){Kessler}, {Steinz}, {Anderegg}, {Clavel},
  {Drechsel}, {Estaria}, {Faelker}, {Riedinger}, {Robson}, {Taylor}, \&
  {Xim{\'e}nez de Ferr{\'a}n}}]{iso96}
{Kessler}, M.~F., {Steinz}, J.~A., {Anderegg}, M.~E., {et~al.} 1996, \aap, 315,
  L27

\bibitem[{{Kholopov} {et~al.}(1992){Kholopov}, {Samus}, {Durlevich},
  {Kazarovets}, {Kireeva}, \& {Tsvetkova}}]{gcvs4}
{Kholopov}, P.~N., {Samus}, N.~N., {Durlevich}, O.~V., {et~al.} 1992, Bulletin
  d'Information du Centre de Donnees Stellaires, 40, 15

\bibitem[{{Kozasa} {et~al.}(1996){Kozasa}, {Dorschner}, {Henning}, \&
  {Stognienko}}]{koz96}
{Kozasa}, T., {Dorschner}, J., {Henning}, T., \& {Stognienko}, R. 1996,
  Astronomy and Astrophysics, 307, 551

\bibitem[{{Kraemer} {et~al.}(2002){Kraemer}, {Sloan}, {Price}, \&
  {Walker}}]{kspw02}
{Kraemer}, K.~E., {Sloan}, G.~C., {Price}, S.~D., \& {Walker}, H.~J. 2002,
  \apjs, 140, 389, \dodoi{10.1086/339708}

\bibitem[{{Lagadec} \& {Zijlstra}(2008)}]{lz08}
{Lagadec}, E., \& {Zijlstra}, A.~A. 2008, \mnras, 390, L59,
  \dodoi{10.1111/j.1745-3933.2008.00535.x}

\bibitem[{{Lagadec} {et~al.}(2007){Lagadec}, {Zijlstra}, {Sloan}, {Matsuura},
  {Wood}, {van Loon}, {Harris}, {Blommaert}, {Hony}, {Groenewegen}, {Feast},
  {Whitelock}, {Menzies}, \& {Cioni}}]{lag07}
{Lagadec}, E., {Zijlstra}, A.~A., {Sloan}, G.~C., {et~al.} 2007, \mnras, 376,
  1270, \dodoi{10.1111/j.1365-2966.2007.11517.x}

\bibitem[{{Leavitt} \& {Pickering}(1912)}]{leavitt+1912}
{Leavitt}, H.~S., \& {Pickering}, E.~C. 1912, Harvard College Observatory
  Circular, 173, 1

\bibitem[{{Leisenring} {et~al.}(2008){Leisenring}, {Kemper}, \&
  {Sloan}}]{leisenringea08}
{Leisenring}, J.~M., {Kemper}, F., \& {Sloan}, G.~C. 2008, \apj, 681, 1557,
  \dodoi{10.1086/588378}

\bibitem[{{Lorenz-Martins} {et~al.}(2001){Lorenz-Martins}, {de Ara{\'u}jo},
  {Codina Land aberry}, {de Almeida}, \& {de Nader}}]{lor01}
{Lorenz-Martins}, S., {de Ara{\'u}jo}, F.~X., {Codina Land aberry}, S.~J., {de
  Almeida}, W.~G., \& {de Nader}, R.~V. 2001, Astronomy and Astrophysics, 367,
  189, \dodoi{10.1051/0004-6361:20000411}

\bibitem[{{Martin} \& {Rogers}(1987)}]{mr87}
{Martin}, P.~G., \& {Rogers}, C. 1987, The Astrophysical Journal, 322, 374,
  \dodoi{10.1086/165736}

\bibitem[{{Massalkhi} {et~al.}(2019){Massalkhi}, {Ag{\'u}ndez}, \&
  {Cernicharo}}]{massalkhiea19}
{Massalkhi}, S., {Ag{\'u}ndez}, M., \& {Cernicharo}, J. 2019, \aap, 628, A62,
  \dodoi{10.1051/0004-6361/201935069}

\bibitem[{{Matsuura} {et~al.}(2002){Matsuura}, {Zijlstra}, {van Loon},
  {Yamamura}, {Markwick}, {Woods}, \& {Waters}}]{matea02}
{Matsuura}, M., {Zijlstra}, A.~A., {van Loon}, J.~T., {et~al.} 2002, The
  Astrophysical Journal, 580, L133, \dodoi{10.1086/345680}

\bibitem[{{Matsuura} {et~al.}(2005){Matsuura}, {Zijlstra}, {van Loon},
  {Yamamura}, {Markwick}, {Whitelock}, {Woods}, {Marshall}, {Feast}, \&
  {Waters}}]{matsuuraea05}
---. 2005, \aap, 434, 691, \dodoi{10.1051/0004-6361:20042305}

\bibitem[{{Matsuura} {et~al.}(2006){Matsuura}, {Wood}, {Sloan}, {Zijlstra},
  {van Loon}, {Groenewegen}, {Blommaert}, {Cioni}, {Feast}, {Habing}, {Hony},
  {Lagadec}, {Loup}, {Menzies}, {Waters}, \& {Whitelock}}]{matsuuraea06}
{Matsuura}, M., {Wood}, P.~R., {Sloan}, G.~C., {et~al.} 2006, \mnras, 371, 415,
  \dodoi{10.1111/j.1365-2966.2006.10664.x}

\bibitem[{{Matsuura} {et~al.}(2007){Matsuura}, {Zijlstra}, {Bernard-Salas},
  {Menzies}, {Sloan}, {Whitelock}, {Wood}, {Cioni}, {Feast}, {Lagadec}, {van
  Loon}, {Groenewegen}, \& {Harris}}]{matsuuraea07}
{Matsuura}, M., {Zijlstra}, A.~A., {Bernard-Salas}, J., {et~al.} 2007, \mnras,
  382, 1889, \dodoi{10.1111/j.1365-2966.2007.12501.x}

\bibitem[{{Matsuura} {et~al.}(2009){Matsuura}, {Barlow}, {Zijlstra},
  {Whitelock}, {Cioni}, {Groenewegen}, {Volk}, {Kemper}, {Kodama}, {Lagadec},
  {Meixner}, {Sloan}, \& {Srinivasan}}]{mbz09}
{Matsuura}, M., {Barlow}, M.~J., {Zijlstra}, A.~A., {et~al.} 2009, \mnras, 396,
  918, \dodoi{10.1111/j.1365-2966.2009.14743.x}

\bibitem[{{Mattei} {et~al.}(1997){Mattei}, {Foster}, {Hurwitz}, {Malatesta},
  {Willson}, \& {Mennessier}}]{matteiea97}
{Mattei}, J.~A., {Foster}, G., {Hurwitz}, L.~A., {et~al.} 1997, in ESA Special
  Publication, Vol. 402, Hipparcos - Venice '97, ed. R.~M. {Bonnet},
  E.~{H{\o}g}, P.~L. {Bernacca}, L.~{Emiliani}, A.~{Blaauw}, C.~{Turon},
  J.~{Kovalevsky}, L.~{Lindegren}, H.~{Hassan}, M.~{Bouffard}, B.~{Strim},
  D.~{Heger}, M.~A.~C. {Perryman}, \& L.~{Woltjer}, 269--274

\bibitem[{{Mattsson} {et~al.}(2008){Mattsson}, {Wahlin}, {H{\"o}fner}, \&
  {Eriksson}}]{mattssonea08}
{Mattsson}, L., {Wahlin}, R., {H{\"o}fner}, S., \& {Eriksson}, K. 2008, \aap,
  484, L5, \dodoi{10.1051/0004-6361:200809689}

\bibitem[{{McCabe}(1982)}]{mccabe82}
{McCabe}, E.~M. 1982, Monthly Notices of the Royal Astronomical Society, 200,
  71, \dodoi{10.1093/mnras/200.1.71}

\bibitem[{{McDonald} {et~al.}(2018){McDonald}, {De Beck}, {Zijlstra}, \&
  {Lagadec}}]{mcdonaldea18}
{McDonald}, I., {De Beck}, E., {Zijlstra}, A.~A., \& {Lagadec}, E. 2018,
  \mnras, 481, 4984, \dodoi{10.1093/mnras/sty2607}

\bibitem[{{McDonald} \& {Trabucchi}(2019)}]{mcdtrabucchi19}
{McDonald}, I., \& {Trabucchi}, M. 2019, \mnras, 484, 4678,
  \dodoi{10.1093/mnras/stz324}

\bibitem[{{Meixner} {et~al.}(2006){Meixner}, {Gordon}, {Indebetouw}, {Hora},
  {Whitney}, {Blum}, {Reach}, {Bernard}, {Meade}, {Babler}, {Engelbracht},
  {For}, {Misselt}, {Vijh}, {Leitherer}, {Cohen}, {Churchwell}, {Boulanger},
  {Frogel}, {Fukui}, {Gallagher}, {Gorjian}, {Harris}, {Kelly}, {Kawamura},
  {Kim}, {Latter}, {Madden}, {Markwick-Kemper}, {Mizuno}, {Mizuno}, {Mould},
  {Nota}, {Oey}, {Olsen}, {Onishi}, {Paladini}, {Panagia}, {Perez-Gonzalez},
  {Shibai}, {Sato}, {Smith}, {Staveley-Smith}, {Tielens}, {Ueta}, {van Dyk},
  {Volk}, {Werner}, \& {Zaritsky}}]{sage06}
{Meixner}, M., {Gordon}, K.~D., {Indebetouw}, R., {et~al.} 2006, \aj, 132,
  2268, \dodoi{10.1086/508185}

\bibitem[{{Micelotta} {et~al.}(2018){Micelotta}, {Matsuura}, \&
  {Sarangi}}]{micelottaea18}
{Micelotta}, E.~R., {Matsuura}, M., \& {Sarangi}, A. 2018, \ssr, 214, 53,
  \dodoi{10.1007/s11214-018-0484-7}

\bibitem[{{Nakashima} {et~al.}(2000){Nakashima}, {Jiang}, {Deguchi},
  {Sadakane}, \& {Nakada}}]{nakashimaea00}
{Nakashima}, J.-i., {Jiang}, B.~W., {Deguchi}, S., {Sadakane}, K., \& {Nakada},
  Y. 2000, Publications of the Astronomical Society of Japan, 52, 275,
  \dodoi{10.1093/pasj/52.2.275}

\bibitem[{{Nanni} {et~al.}(2019){Nanni}, {Groenewegen}, {Aringer}, {Rubele},
  {Bressan}, {van Loon}, {Goldman}, \& {Boyer}}]{nanniea19}
{Nanni}, A., {Groenewegen}, M. A.~T., {Aringer}, B., {et~al.} 2019, \mnras,
  487, 502, \dodoi{10.1093/mnras/stz1255}

\bibitem[{{Nanni} {et~al.}(2017){Nanni}, {Marigo}, {Groenewegen}, {Aringer},
  {Pastorelli}, {Rubele}, {Girardi}, {Bressan}, \& {Bladh}}]{nanniea17}
{Nanni}, A., {Marigo}, P., {Groenewegen}, M.~A.~T., {et~al.} 2017, \memsai, 88,
  393

\bibitem[{{Neugebauer} {et~al.}(1984){Neugebauer}, {Habing}, {van Duinen},
  {Aumann}, {Baud}, {Beichman}, {Beintema}, {Boggess}, {Clegg}, {de Jong},
  {Emerson}, {Gautier}, {Gillett}, {Harris}, {Hauser}, {Houck}, {Jennings},
  {Low}, {Marsden}, {Miley}, {Olnon}, {Pottasch}, {Raimond}, {Rowan-Robinson},
  {Soifer}, {Walker}, {Wesselius}, \& {Young}}]{iras84}
{Neugebauer}, G., {Habing}, H.~J., {van Duinen}, R., {et~al.} 1984, \apjl, 278,
  L1, \dodoi{10.1086/184209}

\bibitem[{{Olnon} {et~al.}(1986){Olnon}, {Raimond}, {Neugebauer}, {van Duinen},
  {Habing}, {Aumann}, {Beintema}, {Boggess}, {Borgman}, {Clegg}, {Gillett},
  {Hauser}, {Houck}, {Jennings}, {de Jong}, {Low}, {Marsden}, {Pottasch},
  {Soifer}, {Walker}, {Emerson}, {Rowan-Robinson}, {Wesselius}, {Baud},
  {Beichman}, {Gautier}, {Harris}, {Miley}, \& {Young}}]{lrs86}
{Olnon}, F.~M., {Raimond}, E., {Neugebauer}, G., {et~al.} 1986, \aaps, 65, 607

\bibitem[{{Payne-Gaposchkin} \& {Gaposchkin}(1938)}]{pgg38}
{Payne-Gaposchkin}, C., \& {Gaposchkin}, S. 1938, {Variable stars}, Vol.~5

\bibitem[{{Piatti}(2012)}]{piatti12}
{Piatti}, A.~E. 2012, Monthly Notices of the Royal Astronomical Society, 422,
  1109, \dodoi{10.1111/j.1365-2966.2012.20684.x}

\bibitem[{{Piatti} \& {Geisler}(2013)}]{piattigeisler13}
{Piatti}, A.~E., \& {Geisler}, D. 2013, The Astronomical Journal, 145, 17,
  \dodoi{10.1088/0004-6256/145/1/17}

\bibitem[{{Rouleau} \& {Martin}(1991)}]{rm91}
{Rouleau}, F., \& {Martin}, P.~G. 1991, \apj, 377, 526, \dodoi{10.1086/170382}

\bibitem[{{Rubele} {et~al.}(2015){Rubele}, {Girardi}, {Kerber}, {Cioni},
  {Piatti}, {Zaggia}, {Bekki}, {Bressan}, {Clementini}, {de Grijs}, {Emerson},
  {Groenewegen}, {Ivanov}, {Marconi}, {Marigo}, {Moretti}, {Ripepi},
  {Subramanian}, {Tatton}, \& {van Loon}}]{rubeleea15}
{Rubele}, S., {Girardi}, L., {Kerber}, L., {et~al.} 2015, \mnras, 449, 639,
  \dodoi{10.1093/mnras/stv141}

\bibitem[{{Samus'} {et~al.}(2017){Samus'}, {Kazarovets}, {Durlevich},
  {Kireeva}, \& {Pastukhova}}]{gcvs17}
{Samus'}, N.~N., {Kazarovets}, E.~V., {Durlevich}, O.~V., {Kireeva}, N.~N., \&
  {Pastukhova}, E.~N. 2017, Astronomy Reports, 61, 80,
  \dodoi{10.1134/S1063772917010085}

\bibitem[{{Sloan}(2017)}]{sloan17}
{Sloan}, G.~C. 2017, in IAU Symposium, Vol. 323, Planetary Nebulae:
  Multi-Wavelength Probes of Stellar and Galactic Evolution, ed. X.~{Liu},
  L.~{Stanghellini}, \& A.~{Karakas}, 121--127,
  \dodoi{10.1017/S1743921317002034}

\bibitem[{{Sloan} {et~al.}(2003{\natexlab{a}}){Sloan}, {Kraemer}, {Goebel}, \&
  {Price}}]{sloan13um}
{Sloan}, G.~C., {Kraemer}, K.~E., {Goebel}, J.~H., \& {Price}, S.~D.
  2003{\natexlab{a}}, \apj, 594, 483, \dodoi{10.1086/376857}

\bibitem[{{Sloan} {et~al.}(2006){Sloan}, {Kraemer}, {Matsuura}, {Wood},
  {Price}, \& {Egan}}]{smcc06}
{Sloan}, G.~C., {Kraemer}, K.~E., {Matsuura}, M., {et~al.} 2006, \apj, 645,
  1118, \dodoi{10.1086/504516}

\bibitem[{{Sloan} {et~al.}(2003{\natexlab{b}}){Sloan}, {Kraemer}, {Price}, \&
  {Shipman}}]{swsatlas}
{Sloan}, G.~C., {Kraemer}, K.~E., {Price}, S.~D., \& {Shipman}, R.~F.
  2003{\natexlab{b}}, \apjs, 147, 379, \dodoi{10.1086/375443}

\bibitem[{{Sloan} {et~al.}(2008){Sloan}, {Kraemer}, {Wood}, {Zijlstra},
  {Bernard-Salas}, {Devost}, \& {Houck}}]{zoo08}
{Sloan}, G.~C., {Kraemer}, K.~E., {Wood}, P.~R., {et~al.} 2008, \apj, 686,
  1056, \dodoi{10.1086/591437}

\bibitem[{{Sloan} {et~al.}(2015){Sloan}, {Lagadec}, {Kraemer}, {Boyer},
  {Srinivasan}, {McDonald}, \& {Zijlstra}}]{slk15}
{Sloan}, G.~C., {Lagadec}, E., {Kraemer}, K.~E., {et~al.} 2015, in Astronomical
  Society of the Pacific Conference Series, Vol. 497, Why Galaxies Care about
  AGB Stars III: A Closer Look in Space and Time, ed. F.~{Kerschbaum}, R.~F.
  {Wing}, \& J.~{Hron}, 429.
\newblock \doarXiv{1412.1845}

\bibitem[{{Sloan} {et~al.}(1996){Sloan}, {Levan}, \&
  {Little-Marenin}}]{sloanea96}
{Sloan}, G.~C., {Levan}, P.~D., \& {Little-Marenin}, I.~R. 1996, \apj, 463,
  310, \dodoi{10.1086/177243}

\bibitem[{{Sloan} {et~al.}(1998){Sloan}, {Little-Marenin}, \& {Price}}]{slmp98}
{Sloan}, G.~C., {Little-Marenin}, I.~R., \& {Price}, S.~D. 1998, \aj, 115, 809,
  \dodoi{10.1086/300205}

\bibitem[{{Sloan} {et~al.}(2012){Sloan}, {Matsuura}, {Lagadec}, {van Loon},
  {Kraemer}, {McDonald}, {Groenewegen}, {Wood}, {Bernard-Salas}, \&
  {Zijlstra}}]{sml12}
{Sloan}, G.~C., {Matsuura}, M., {Lagadec}, E., {et~al.} 2012, \apj, 752, 140,
  \dodoi{10.1088/0004-637X/752/2/140}

\bibitem[{{Sloan} {et~al.}(2016){Sloan}, {Kraemer}, {McDonald}, {Groenewegen},
  {Wood}, {Zijlstra}, {Lagadec}, {Boyer}, {Kemper}, {Matsuura}, {Sahai},
  {Sargent}, {Srinivasan}, {van Loon}, \& {Volk}}]{mcc16}
{Sloan}, G.~C., {Kraemer}, K.~E., {McDonald}, I., {et~al.} 2016, \apj, 826, 44,
  \dodoi{10.3847/0004-637X/826/1/44}

\bibitem[{{Soszy{\'n}ski} {et~al.}(2013){Soszy{\'n}ski}, {Wood}, \&
  {Udalski}}]{swu13}
{Soszy{\'n}ski}, I., {Wood}, P.~R., \& {Udalski}, A. 2013, \apj, 779, 167,
  \dodoi{10.1088/0004-637X/779/2/167}

\bibitem[{{Soszy{\'n}ski} {et~al.}(2009){Soszy{\'n}ski}, {Udalski},
  {Szyma{\'n}ski}, {Kubiak}, {Pietrzy{\'n}ski}, {Wyrzykowski}, {Szewczyk},
  {Ulaczyk}, \& {Poleski}}]{sus09}
{Soszy{\'n}ski}, I., {Udalski}, A., {Szyma{\'n}ski}, M.~K., {et~al.} 2009,
  \actaa, 59, 239.
\newblock \doarXiv{0910.1354}

\bibitem[{{Soszy{\'n}ski} {et~al.}(2011){Soszy{\'n}ski}, {Udalski},
  {Szyma{\'n}ski}, {Kubiak}, {Pietrzy{\'n}ski}, {Wyrzykowski}, {Ulaczyk},
  {Poleski}, {Koz{\l}owski}, \& {Pietrukowicz}}]{sus11}
---. 2011, \actaa, 61, 217.
\newblock \doarXiv{1109.1143}

\bibitem[{{Speck} {et~al.}(2006){Speck}, {Cami}, {Markwick-Kemper},
  {Leisenring}, {Szczerba}, {Dijkstra}, {Van Dyk}, \& {Meixner}}]{speckea06}
{Speck}, A.~K., {Cami}, J., {Markwick-Kemper}, C., {et~al.} 2006, \apj, 650,
  892, \dodoi{10.1086/507178}

\bibitem[{{Udalski} {et~al.}(1992){Udalski}, {Szymanski}, {Kaluzny}, {Kubiak},
  \& {Mateo}}]{ogle92}
{Udalski}, A., {Szymanski}, M., {Kaluzny}, J., {Kubiak}, M., \& {Mateo}, M.
  1992, \actaa, 42, 253

\bibitem[{{Uttenthaler} {et~al.}(2019){Uttenthaler}, {McDonald}, {Bernhard},
  {Cristallo}, \& {Gobrecht}}]{uttea19}
{Uttenthaler}, S., {McDonald}, I., {Bernhard}, K., {Cristallo}, S., \&
  {Gobrecht}, D. 2019, \aap, 622, A120, \dodoi{10.1051/0004-6361/201833794}

\bibitem[{{van Loon} {et~al.}(2006){van Loon}, {Marshall}, {Cohen}, {Matsuura},
  {Wood}, {Yamamura}, \& {Zijlstra}}]{vlea06}
{van Loon}, J.~T., {Marshall}, J.~R., {Cohen}, M., {et~al.} 2006, Astronomy and
  Astrophysics, 447, 971, \dodoi{10.1051/0004-6361:20054222}

\bibitem[{{van Loon} {et~al.}(1999){van Loon}, {Zijlstra}, \&
  {Groenewegen}}]{vlea99}
{van Loon}, J.~T., {Zijlstra}, A.~A., \& {Groenewegen}, M.~A.~T. 1999,
  Astronomy and Astrophysics, 346, 805.
\newblock \doarXiv{astro-ph/9902284}

\bibitem[{{Wallerstein} \& {Knapp}(1998)}]{wk98}
{Wallerstein}, G., \& {Knapp}, G.~R. 1998, \araa, 36, 369,
  \dodoi{10.1146/annurev.astro.36.1.369}

\bibitem[{{Werner} {et~al.}(2004){Werner}, {Roellig}, {Low}, {Rieke}, {Rieke},
  {Hoffmann}, {Young}, {Houck}, {Brandl}, {Fazio}, {Hora}, {Gehrz}, {Helou},
  {Soifer}, {Stauffer}, {Keene}, {Eisenhardt}, {Gallagher}, {Gautier}, {Irace},
  {Lawrence}, {Simmons}, {Van Cleve}, {Jura}, {Wright}, \&
  {Cruikshank}}]{spitzer04}
{Werner}, M.~W., {Roellig}, T.~L., {Low}, F.~J., {et~al.} 2004, \apjs, 154, 1,
  \dodoi{10.1086/422992}

\bibitem[{{Whitelock} {et~al.}(2006){Whitelock}, {Feast}, {Marang}, \&
  {Groenewegen}}]{wfm06}
{Whitelock}, P.~A., {Feast}, M.~W., {Marang}, F., \& {Groenewegen}, M.~A.~T.
  2006, \mnras, 369, 751, \dodoi{10.1111/j.1365-2966.2006.10322.x}

\bibitem[{{Whitelock} {et~al.}(2003){Whitelock}, {Feast}, {van Loon}, \&
  {Zijlstra}}]{whitelockea03}
{Whitelock}, P.~A., {Feast}, M.~W., {van Loon}, J.~T., \& {Zijlstra}, A.~A.
  2003, \mnras, 342, 86, \dodoi{10.1046/j.1365-8711.2003.06514.x}

\bibitem[{{Wood} \& {Sebo}(1996)}]{ws96}
{Wood}, P.~R., \& {Sebo}, K.~M. 1996, \mnras, 282, 958

\bibitem[{{Wood} {et~al.}(1999){Wood}, {Alcock}, {Allsman}, {Alves}, {Axelrod},
  {Becker}, {Bennett}, {Cook}, {Drake}, {Freeman}, {Griest}, {King}, {Lehner},
  {Marshall}, {Minniti}, {Peterson}, {Pratt}, {Quinn}, {Stubbs}, {Sutherland},
  {Tomaney}, {Vandehei}, \& {Welch}}]{woodea99}
{Wood}, P.~R., {Alcock}, C., {Allsman}, R.~A., {et~al.} 1999, in IAU Symposium,
  Vol. 191, Asymptotic Giant Branch Stars, ed. T.~{Le Bertre}, A.~{Lebre}, \&
  C.~{Waelkens}, 151

\bibitem[{{Wright} {et~al.}(2010){Wright}, {Eisenhardt}, {Mainzer}, {Ressler},
  {Cutri}, {Jarrett}, {Kirkpatrick}, {Padgett}, {McMillan}, {Skrutskie},
  {Stanford}, {Cohen}, {Walker}, {Mather}, {Leisawitz}, {Gautier}, {McLean},
  {Benford}, {Lonsdale}, {Blain}, {Mendez}, {Irace}, {Duval}, {Liu}, {Royer},
  {Heinrichsen}, {Howard}, {Shannon}, {Kendall}, {Walsh}, {Larsen}, {Cardon},
  {Schick}, {Schwalm}, {Abid}, {Fabinsky}, {Naes}, \& {Tsai}}]{wise10}
{Wright}, E.~L., {Eisenhardt}, P.~R.~M., {Mainzer}, A.~K., {et~al.} 2010, \aj,
  140, 1868, \dodoi{10.1088/0004-6256/140/6/1868}

\bibitem[{Young {et~al.}(2012)Young, Becklin, Marcum, Roellig, Buizer, Herter,
  Güsten, Dunham, Temi, Andersson, Backman, Burgdorf, Caroff, Casey, Davidson,
  Erickson, Gehrz, Harper, Harvey, Helton, Horner, Howard, Klein, Krabbe,
  McLean, Meyer, Miles, Morris, Reach, Rho, Richter, Roeser, Sandell, Sankrit,
  Savage, Smith, Shuping, Vacca, Vaillancourt, Wolf, \& Zinnecker}]{sofia}
Young, E.~T., Becklin, E.~E., Marcum, P.~M., {et~al.} 2012, The Astrophysical
  Journal Letters, 749, L17.
\newblock \url{http://stacks.iop.org/2041-8205/749/i=2/a=L17}

\bibitem[{{Zijlstra} {et~al.}(2006){Zijlstra}, {Matsuura}, {Wood}, {Sloan},
  {Lagadec}, {van Loon}, {Groenewegen}, {Feast}, {Menzies}, {Whitelock},
  {Blommaert}, {Cioni}, {Habing}, {Hony}, {Loup}, \& {Waters}}]{zijlstraea06}
{Zijlstra}, A.~A., {Matsuura}, M., {Wood}, P.~R., {et~al.} 2006, \mnras, 370,
  1961, \dodoi{10.1111/j.1365-2966.2006.10623.x}

\bibitem[{{Zubko} {et~al.}(2004){Zubko}, {Dwek}, \& {Arendt}}]{zubkoea04}
{Zubko}, V., {Dwek}, E., \& {Arendt}, R.~G. 2004, \apjs, 152, 211,
  \dodoi{10.1086/382351}

\end{thebibliography}

\end{document}